\documentclass[runningheads]{llncs}
\usepackage{url}

\usepackage{graphicx}
\usepackage{adjustbox}
\usepackage{multirow}
\title{Proactive Provenance Policies for Automatic Cryptographic Data Centric Security}
\author{Shamaria Engram \and Tyler Kaczmarek \and Alice Lee \and David Bigelow}
\institute{
	MIT Lincoln Laboratory \\
	\email{\{shamaria.engram, tyler.kaczmarek, alice.lee, dbigelow\}@ll.mit.edu}
}
\titlerunning{Proactive Provenance Policies for ACDC}
\begin{document}
\maketitle
\begin{abstract}
Data provenance analysis has been used as an assistive measure for ensuring system integrity. However, such techniques are typically reactive approaches to identify the root cause of an attack in its aftermath. This is in part due to fact that the collection of provenance metadata often results in a deluge of information that cannot easily be queried and analyzed in real time. This paper presents an approach for proactively reasoning about provenance metadata within the Automatic Cryptographic Data Centric (ACDC) security architecture, a new security infrastructure in which all data interactions are considered at a coarse granularity, similar to the Function as a Service model. At this scale, we have found that data interactions are manageable for the proactive specification and evaluation of {\em provenance policies}---constraints placed on provenance metadata to prevent the consumption of untrusted data. This paper provides a model for proactively evaluating provenance metadata in the ACDC paradigm as well as a case study of an electronic voting scheme to demonstrate the applicability of ACDC and the provenance policies needed to ensure data integrity.
\end{abstract}

\let\thefootnote\relax\footnote{DISTRIBUTION STATEMENT A. Approved for public release. Distribution is unlimited.

This material is based upon work supported by the Under Secretary of Defense for Research and Engineering under Air Force Contract No. FA8702-15-D-0001. Any opinions, findings, conclusions or recommendations expressed in this material are those of the author(s) and do not necessarily reflect the views of the Under Secretary of Defense for Research and Engineering.}

\section{Introduction}\label{sec:intro}
Data provenance provides a comprehensive history of data and the manipulations it has underwent from its inception to its latest state. Analysis of this history can provide significant insight into a datum's integrity and authenticity for forensic analysts and security administrators. However, due to the mass of data being produced in computing environments, manual analysis of provenance metadata is a daunting task. Automated provenance analysis techniques exist but generally provide a reactive evaluation in the aftermath of a security incident (e.g.,\cite{lemay2017automated}).

This retrospective approach to data provenance analysis has proven valuable in several security contexts (e.g., diagnosing an attacker's point of entry to a system). Nevertheless, given the ubiquity of online services, many of which operate in an outsourced distributed environment, there is a need for a proactive approach to data provenance analysis. Proactively evaluating a datum's provenance record before consumption is especially applicable to operations within cloud environments, where end users, who outsource their data to be processed by cloud applications, should have some level of assurance about their data's integrity. Runtime analysis of whole-system provenance has recently gained attention in the literature but does so at a fine-grained level, which does not translate cleanly to a distributed system~\cite{pasquier2018runtime}. 

The ability to proactively specify properties of provenance metadata, to aid in security enforcement decisions, can have a significant impact on a distributed environment's overall security posture. This paper presents an approach for proactively reasoning about provenance metadata within the Automatic Cryptographic Data Centric (ACDC) security architecture, a distributed architecture that upends the current system-centric paradigm by taking a data-centric approach to security. Rather than protecting systems that store data, ACDC puts the focus directly on protecting data itself both at rest and in motion while simultaneously ensuring that data is used in only authorized and auditable ways. Data protection goals include confidentiality, integrity, and availability throughout all uses of the data, including not only storage and transmission but also sharing and computation, on devices and networks that may be partially compromised. 

ACDC allows application developers to proactively express policies over provenance metadata to be enforced before data is consumed by an individual process. We call such policies {\em provenance policies}. ACDC can prevent the consumption of untrusted data by providing the following capabilities: 1)~secure packaging of data with associated integrity and confidentiality policies at the network's edge, 2)~enforcement of integrity and confidentiality policies throughout the data's entire lifespan, and 3)~a thorough record of data provenance to account for every manipulation. To the best of our knowledge, this paper presents the first effort to provide a proactive approach for data provenance evaluation within a data-centric security architecture.

Our core contributions are as follows:

\begin{enumerate}
	\item We introduce the ACDC architecture for data-centric security (Section~\ref{sec:arch}),
	\item We describe a formal approach for reasoning about provenance policies proactively based on a mathematical semantics of provenance metadata (Section~\ref{sec: preliminaries}), and
	\item We demonstrate the applicability of ACDC and proactive provenance policy evaluation by providing a case study of an end-to-end, coercion-resistant voting system (Section~\ref{sec:cs}).
\end{enumerate}

Section~\ref{sec: background} provides a summary of related work and Section~\ref{sec:conc} concludes and provides directions for future work.

\newcommand\todo[1]{\textcolor{red}{TODO: {#1}}}
\newcommand\tbd[1]{\textcolor{red}{TODO: {#1}}}
\newcommand\subsubsubsection[1]{{\bf #1}\par}
\newcommand\pterm[1]{{\tt#1}}

\makeatletter
\renewcommand\paragraph{\@startsection{paragraph}{4}{\z@}%
	{\parskip}
	{-1em}%
	{\normalfont\normalsize\bfseries}}
\makeatother

\section{The ACDC FaaS Paradigm}\label{sec:arch}
\begin{figure}[t]
	\centering \includegraphics[trim=0 2.5cm 0 2.7cm,clip,width=.95\linewidth]{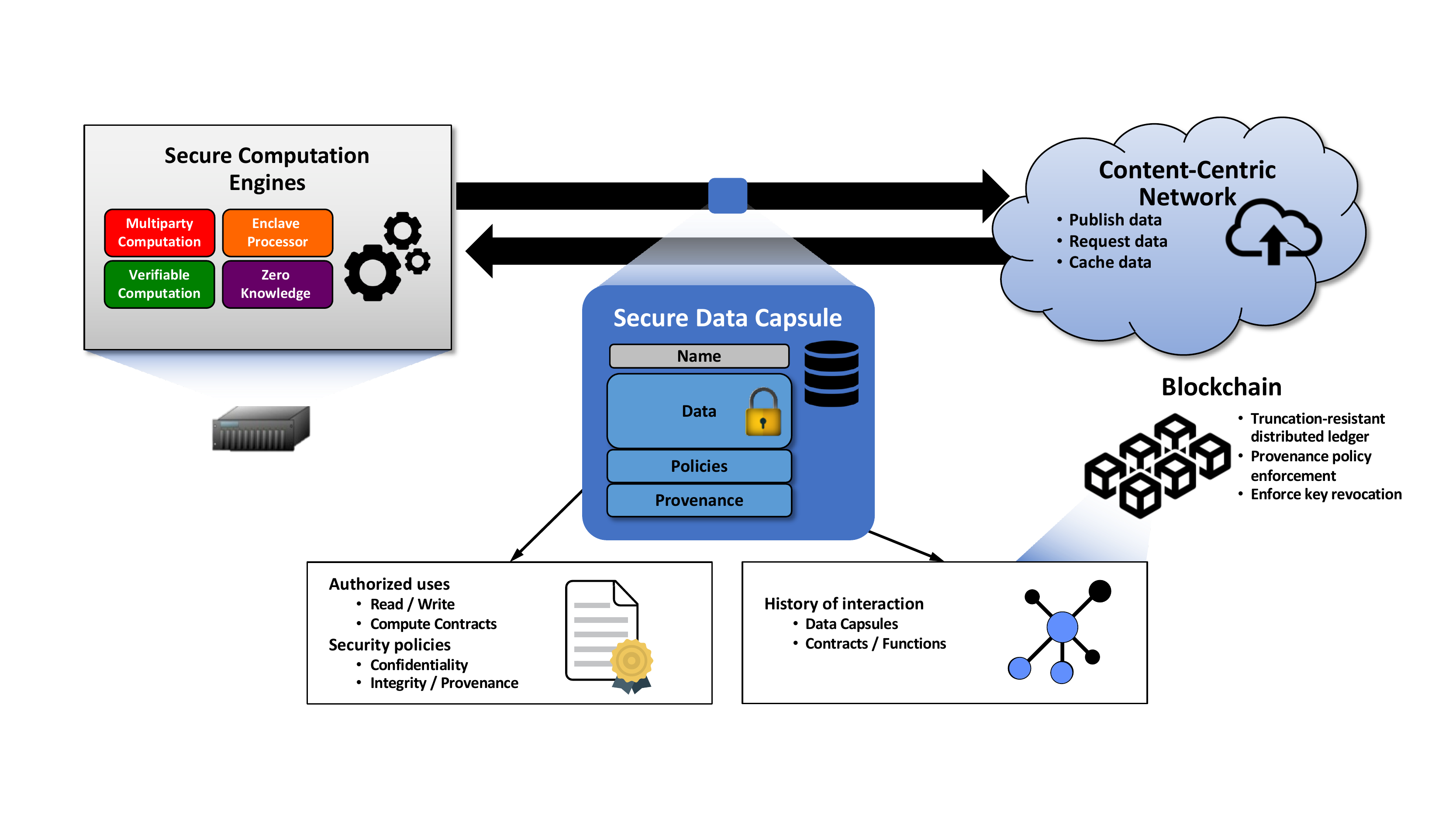}
	\caption{ACDC Core Component Architecture}\label{fig:acdc-arch}
\end{figure}
This section introduces the Automatic Cryptographic Data-Centric (ACDC) security paradigm and describes each of the components that make up an ACDC network. As shown in Figure~\ref{fig:acdc-arch}, ACDC puts all data into named, secure data capsules, where each capsule is associated with an owner. These capsules contain cryptographically enforced access-control policies that define who can access and use the capsules' associated data. Each capsule also contains its provenance as captured within the ACDC system, allowing authorized parties to assess a capsule's integrity before acting upon it. ACDC provides flexibility to data owners by allowing them to  1)~cryptographically authorize functions to run on their data, and 2)~specify which secure computation techniques are allowed to process their data (e.g, multiparty computation (MPC) or secure enclaves), which enables data owners to consider the tradoffs between security, functionality, and performance. These capabilities allow mutually distrusting data owners to securely collaborate and share their data in a controlled environment. Lastly, ACDC uses content-centric networking (CCN)~\cite{Jacobson:2009:NNC:1658939.1658941} to route and transmit data capsules by their name rather than by the systems storing such data, thus enabling capsules' cryptographic mechanisms to protect data wherever capsules go on the network.

An instance of an ACDC network (closed or Internet-wide) consists of the following components:

\paragraph{\bf Nodes}
ACDC nodes may be a set of dedicated servers each running ACDC software.  Each node may also have a set of supporting servers that provide data for specific ACDC functionality using unspecified (back-end) protocols. In general, all ACDC nodes use a common ACDC core library. The library itself makes no distinction based on the node type, though the capabilities of an individual node can dictate many different types.
\paragraph{\bf Data Capsules}
As previously mentioned, all data is stored in named, secure capsules. All capsules are digitally signed for authenticity and integrity, and the internal data of each capsule is encrypted for confidentiality. Each data capsule may contain an optional output confidentiality policy, which defines the confidentiality restrictions imposed on any data derived from its data.

\paragraph{\bf Capsule Storage}
ACDC stores data capsules persistently, allowing nodes to publish new capsules, fetch existing capsules, and delete capsules. All capsules are named according to a CCN-compatible ACDC naming scheme.
\paragraph{\bf Function as a Service}
FaaS allows nodes to perform (or serve) one or more functions in a query/response model. In general, FaaS is expected to use the same naming schemes as capsule storage, such that any request can be static (Capsule Storage) or dynamic (FaaS).
\paragraph{\bf Secure Execution Environments}
ACDC provides environments for secure function execution (e.g., secure enclaves such as Intel SGX or MPC).
\paragraph{\bf Keys}
ACDC uses cryptographic keys for confidentiality, integrity, and authenticity.
\paragraph{\bf Policies}
ACDC has two types of policies: 1) confidentiality policies, and 2) integrity policies (i.e., provenance policies). The confidentiality policies are attribute-based encryption policies~\cite{chase2007multi} that define the attributes needed to decrypt a data capsule and thus cryptographically enforce access control. Attributes are terms that may refer to a principal's characteristics (e.g., a role or identity) or proof of taking an action (e.g., validating a capsule's provenance). Provenance policies define a capsule's expected provenance and should be checked before a capsule is used as input to a function (discussed at length in Section~\ref{sec: preliminaries}).

\paragraph{\bf Contracts}
Contracts define functions and give restrictions, limiting nodes to perform computations on data capsules under a given set of conditions. For example, a contract may restrict who can perform computations, require provenance checks via a provenance policy (detailed in following sections), or require key revocation checks.

All contracts are expected to provide an output confidentiality policy, which defines confidentiality restrictions to impose on the output data of the function. However, each function argument may have its own output confidentiality policy, in which case the policies must be composed, thereby accumulating all the restrictions from each policy (i.e., the contract and each function argument's output confidentiality policy).

\section{ACDC Provenance Model}\label{sec: preliminaries}

To reason about provenance within an ACDC architecture, we follow the W3C PROV Data Model~\cite{w3c-prov-dm} in characterizing the elements of the model into 3 main types: entities, activities, and agents. We further refine the model by extending the entity type to contain 3 subtypes and the agent type to contain 2 subtypes. An entity can be either a \emph{key entity}, a \emph{contract entity}, or a \emph{data entity} and an agent can be either an \emph{account agent} or a \emph{node agent}.

\begin{table}[t]
	\centering
	\resizebox{\textwidth}{!}{
		\begin{tabular}{@{\extracolsep{8pt}}| c || c | c | c |}
			\hline
			{\bf Relation} & {\bf Source} & {\bf Destination} & {\bf Meaning}\\
			\hline
			\multirow{2}{*}{WasAttributedTo} & \multirow{2}{*}{entity (any subtype)} & node agent & The entity was created by \\
			&&&execution on the node agent.\\
			\cline{3-4}
			& & account agent & The entity was sealed \\
			&&&under the account agent's key(s).\\
			\hline
			\hline
			\multirow{3}{*}{WasDerivedFrom} & \multirow{3}{*}{entity (any subtype)} &  contract entity & The entity was created based \\
			&&&on rules specified in the contract.\\
			\cline{3-4}
			&& data entity & The entity is dependent \\
			&&&on the data entity. \\
			\cline{3-4}
			&& key entity & The key entity was needed to either wrap\\
			&&& the source entity or unwrap an input entity.\\
			\hline
			\hline
			\multirow{3}{*}{Used} & \multirow{3}{*}{activity} & contract entity & The contract entity defined\\
			&&& the activity's execution.\\
			\cline{3-4}
			&& data entity & The data entity was input \\
			&&&to the activity.\\
			\cline{3-4}
			&& key entity & The activity performed some cryptographic\\
			&&& function using the key entity.\\
			\hline
			\hline
			ActedOnBehalfOf & node agent & account agent & The node agent performed a computation\\
			&&& on behalf of the account agent.\\
			\hline
			\hline
			WasAssociatedWith & activity & node agent & The activity describing the computation\\
			&&& was performed by the node agent.\\
			\hline
	\end{tabular}}
	\vspace{0.05cm}
	\caption{The effect of the additional subtypes on provenance relations introduced by ACDC to the PROV data model.}\label{fig:extend}
\end{table}

Key entities represent cryptographic keys belonging to an agent, contract entities represent ACDC contracts, and data entities represent all other types of data. Account agents represent the users in a computing environment and node agents represent a secure execution environment (e.g., an sgx enclave). Activities represent a computation that uses, manipulates, or generates entities. Node agents act on behalf of account agents; conversely, account agents \emph{cannot} act on behalf of node agents. Because node agents represent environments where computations are performed, activities can only be associated with node agents. Table~\ref{fig:extend} summarizes the valid types for provenance relations affected by our additional subtypes.



\begin{figure}[t]
	\centering
	\scalebox{0.85}{\includegraphics[trim =  3.5cm 0 0 2cm,clip, width=\linewidth]{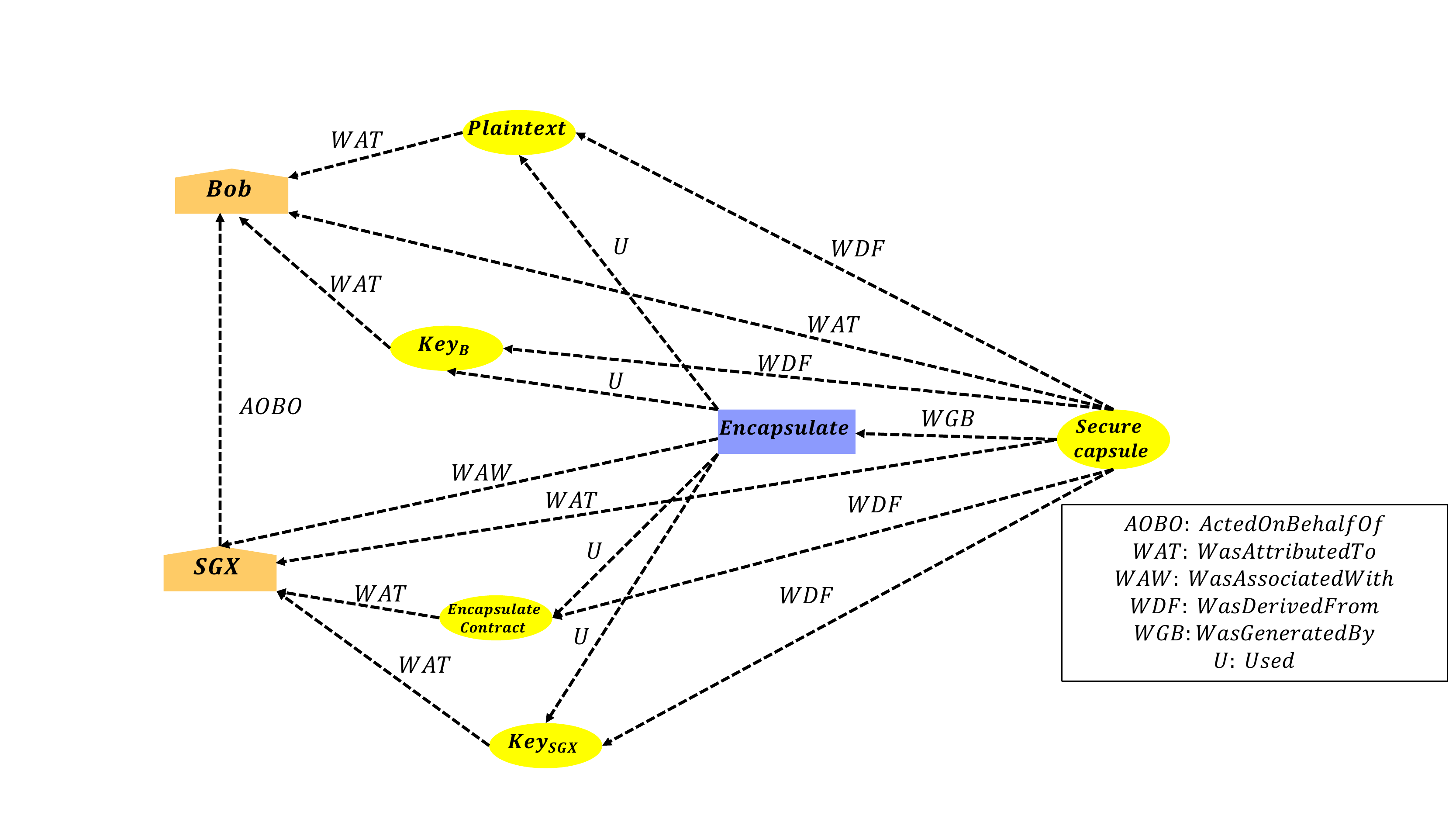}}
	\caption{A provenance graph of a user who has encapsulated some data}\label{fig: encrypt}
\end{figure}
To illustrate this new distinction between entity and agent subtypes, consider the provenance of a scenario in which a user has introduced some data into the ACDC ecosystem at the network's edge, shown in Figure~\ref{fig: encrypt}. To introduce this data, the data must be encapsulated because all data in ACDC is stored in secure capsules. The sgx enclave is a node agent which acts on behalf of Bob who is an account agent. The encapsulate computation is an activity associated with the sgx enclave. The plaintext is a data entity, the encapsulate contract is a contract entity specifying how the function should input and output entities, $Key_{SGX}$ is a key entity attributed to the sgx enclave for secure computation, and $Key_{B}$ is a key entity attributed to account agent Bob. The secure capsule is a data entity generated by the encapsulate activity, derived from the contract, key, and data entities, and is attributed to account agent Bob.

%

To reason about the provenance of a distributed ACDC environment, we specify the environment at a high level of abstraction as a 6-tuple $D = (\mathcal{E}_{k},\mathcal{E}_{c}, \mathcal{E}_{d},$ \linebreak $G_{n}, G_{a},\mathcal{A})$, where $\mathcal{E}_{k}$ is a finite set of key entities ranged over by metavariable $\varepsilon_k$, $\mathcal{E}_{c}$ is a finite set of contract entities ranged over by metavariable $\varepsilon_c$, $\mathcal{E}_{d}$ is a finite set of data entities ranged over by metavariable $\varepsilon_d$, $G_{n}$ is a finite set of node agents ranged over by metavriable $g_n$, $G_{a}$ is a finite set of account agents ranged over by metavariable $g_a$, and $\mathcal{A}$ is a finite set of activities ranged over by metavariable $a$.

The set of all possible entities $\mathcal{E} =  \mathcal{E}_{k}\cup \mathcal{E}_{c} \cup \mathcal{E}_{d} $ is the union of all entity subtypes, and the set of all possible agents $G = G_n\cup G_a$ is the union of all agent subtypes. Because provenance is represented by a labeled, directed acyclic graph,  $V = \mathcal{E} \cup G \cup \mathcal{A}$ denotes the set of all possible vertices, $E \subset V \times V$ denotes the set of all possible edges, $L$ denotes the set of all possible labels (relations) and is the union of all relations, and $L^{E}$ denotes the set of all possible graph labeling functions where $l: E \rightarrow L$ is a function that inputs an edge and outputs the label corresponding to that edge, indicating the causal relationship between the source and destination nodes. 


The set of all provenance graphs of a distributed environment $D$ is denoted by $2^{V}\times 2^{E}\times L^{E}$. A provenance policy is a predicate ${P: 2^{V} \times 2^{E} \times L^{E} \rightarrow \{true,false\}}$. ACDC provenance policies determine whether a particular subgraph is contained in the provenance graph under consideration. It is not always the case that the entire provenance record for a distributed environment be evaluated against a policy. For example, a provenance policy can be evaluated at runtime to ensure that data was generated via the expected pathways before using the data as input for a computation. In this case, a contract will specify a provenance policy to be evaluated over the function's inputs; therefore, only the provenance associated with the input data is relevant for policy evaluation, making it unnecessary and inefficient to evaluate the policy on the entire provenance record. Consequently, for each distributed environment there is a one-to-many relationship between the distributed environment and the number of provenance graphs it contains. In this paper, we refer to an \emph{event} as a provenance subgraph containing an activity with all of its immediate input and output entities along with their attributions. In a larger distributed environment, Figure~\ref{fig: encrypt} would be considered the $Encapsulate$ event.

Provenance policies are specified as boolean predicates so that large, complex policies can be composed from simpler policies. For example, let's consider a scenario where Bob would like to use his secure capsule in a computation, but would like to verify that his secure capsule was properly encapsulated (i.e., encapsulated with only his data and key). A policy for this situation might ensure that: (1) the encapsulate function used Bob's data and key, (2) if the encapsulate function used any data and cryptographic keys, then they can only be Bob's data and key or the node acting on Bob's behalf key, (3) the secure capsule is only derived from Bob's key and plaintext data and no other account agent's key and data, and (4) the secure capsule was computed using the encapsulate contract. To note the importance of precise policy specification, it may not be easy to distinguish the difference between the informal specification of concern (1) and concern (2). Concern (1) only ensures that the encapsulate function used Bob's data and key but does not preclude the function from using any one else's data and key. The second concern ensures that if the encapsulate function used any data or cryptographic keys, then the data and keys can only belong to Bob or the node acting on Bob's behalf. Formally, given a provenance graph $(V',E',l') \in 2^{V}\times 2^{E} \times L^{E}$, Bob can specify the following policies:
\begin{flushleft}
	\begin{tabular}{lcp{1in}}
		\rule{0pt}{3ex}\small $P_{1}(V',E',l')$ & \small $\iff$ & \small${\exists \varepsilon_k\in V': (Encapsulate, \varepsilon_k)\in E' \land l'(Encapsulate, \varepsilon_k) = Used}$,\\
		\rule{0pt}{3ex}\small $P_2(V',E',l')$ & \small $\iff$ & \small${\exists \varepsilon_d\in V': (Encapsulate, \varepsilon_d)\in E' \land l'(Encapsulate, \varepsilon_d) = Used}$,\\
		\rule{0pt}{3ex}\small $P_3(V',E',l')$ &\small$\iff$ & \small${\forall \varepsilon_k\in V': ((Encapsulate, \varepsilon_k) \in E'\land l'(Encapsulate, \varepsilon_k) = Used})\linebreak{\Rightarrow (((\varepsilon_k,Bob)\in E'} {\land~l'(\varepsilon_k, Bob) = WasAttributedTo)}\linebreak {\lor (\exists g_n\in V': ((\varepsilon_k,g_n)\in E'\land~l'(\varepsilon_k,g_n) = WasAttributedTo)}\linebreak{ \land ((g_n,Bob)\in E'\land~l'(g_n, Bob) = ActedOnBehalfOf)))}$,\\
		\rule{0pt}{3ex}\small $P_4(V',E',l')$ &\small$\iff$ & \small${\forall \varepsilon_d\in V': ((Encapsulate, \varepsilon_d) \in E'\land l'(Encapsulate, \varepsilon_d) = Used})\linebreak{\Rightarrow ((\varepsilon_d,Bob)\in E'} {\land~l'(\varepsilon_d, Bob) = WasAttributedTo)}$,\\
		\rule{0pt}{3ex}\small $P_5(V',E',l')$ &\small $\iff$ & \small ${\exists \varepsilon_d \in V': (SecureCapsule,\varepsilon_d) \in E'}\linebreak{\land~l'(SecureCapsule,\varepsilon_d) = WasDerivedFrom}$,\\
		\rule{0pt}{3ex}\small $P_6(V',E',l')$ &\small $\iff$ & \small ${\exists \varepsilon_k\in V': (SecureCapsule,\varepsilon_k) \in E'}\linebreak{\land~l'(SecureCapsule,\varepsilon_k) = WasDerivedFrom}$,\\
		\rule{0pt}{3ex}\small $P_7(V',E',l')$ &\small$\iff$ & \small${\forall \varepsilon_k \in V': ((SecureCapsule, \varepsilon_k) \in E'}\linebreak{\land~l'(SecureCapsule, \varepsilon_k) = WasDerivedFrom)}\linebreak{\Rightarrow (((\varepsilon_k,Bob)\in E'\land~l'(\varepsilon_k, Bob) = WasAttributedTo)}\linebreak{\lor (\exists g_n\in V': ((\varepsilon_k,g_n)\in E'\linebreak \land~l'(\varepsilon_k,g_n) = WasAttributedTo)}\linebreak {\land ((g_n,Bob)\in E'\land~l'(g_n, Bob) = ActedOnBehalfOf)))}$, \\
		\rule{0pt}{3ex}\small $P_8(V',E',l')$ &\small$\iff$ & \small${\forall \varepsilon_d \in V': ((SecureCapsule, \varepsilon_d) \in E'}\linebreak{\land~l'(SecureCapsule, \varepsilon_d) = WasDerivedFrom)}\linebreak{\Rightarrow ((\varepsilon_d,Bob)\in E'\land~l'(\varepsilon_d, Bob) = WasAttributedTo)}$,\\
		\rule{0pt}{3ex}\small $P_9(V',E',l')$ &\small $\iff$ &\small${(SecureCapsule,EncapsulateContract) \in E'}\linebreak {\land~l'(SecureCapsule,EncapsulateContract)= WasDerivedFrom}$.
	\end{tabular}
\end{flushleft}

The overall provenance policy can be composed as the conjunction of policies $P_{1} - P_{9}$. Specifying policies in this way allows analyst to reason about small, simple policies. Logical connectives can then be used to compose these simple policies into larger, more complex policies.

\section{A Case Study on Detecting Voter Fraud in E-voting}\label{sec:cs}
This section presents a case study of an e-voting scenario within an ACDC architecture and provenance policies that may prevent illegal ballots from being cast. As recent voting elections have been under scrutiny by both the media and general public~\cite{cassidy2018voting}, we believe that ACDC equipped voting machines can provide significant benefits and increase public confidence in the integrity of voting elections. 

\begin{table}[!th]
	\centering
	\includegraphics[trim =  0.5cm 1.6cm 2.75cm 0.5cm,clip, width=\linewidth]{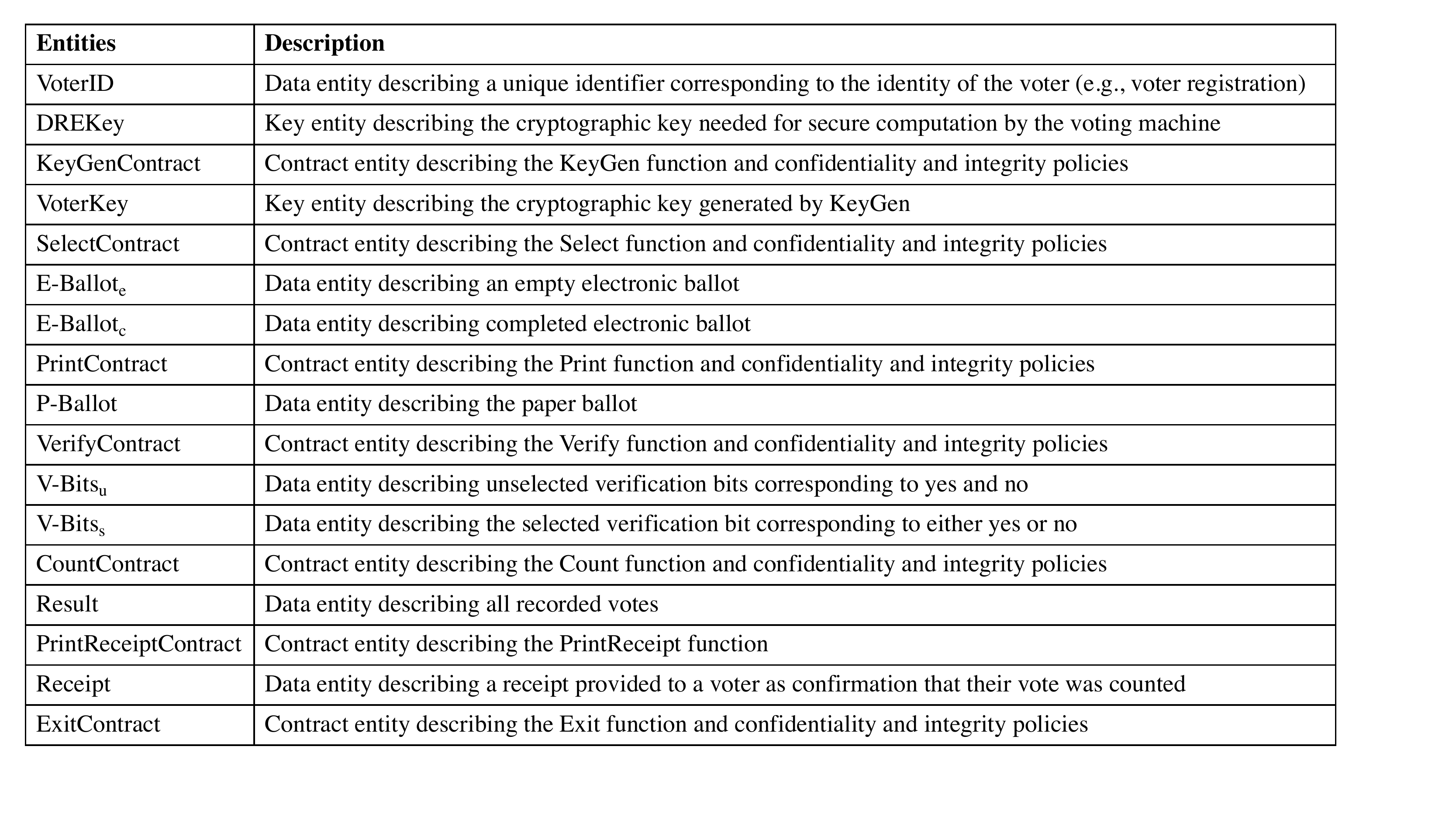}
	\caption{Entities in an ACDC E-voting environment}\label{tab: voter-entities}
	\includegraphics[trim =  1cm 9.5cm 2.5cm 2cm,clip, width=\linewidth]{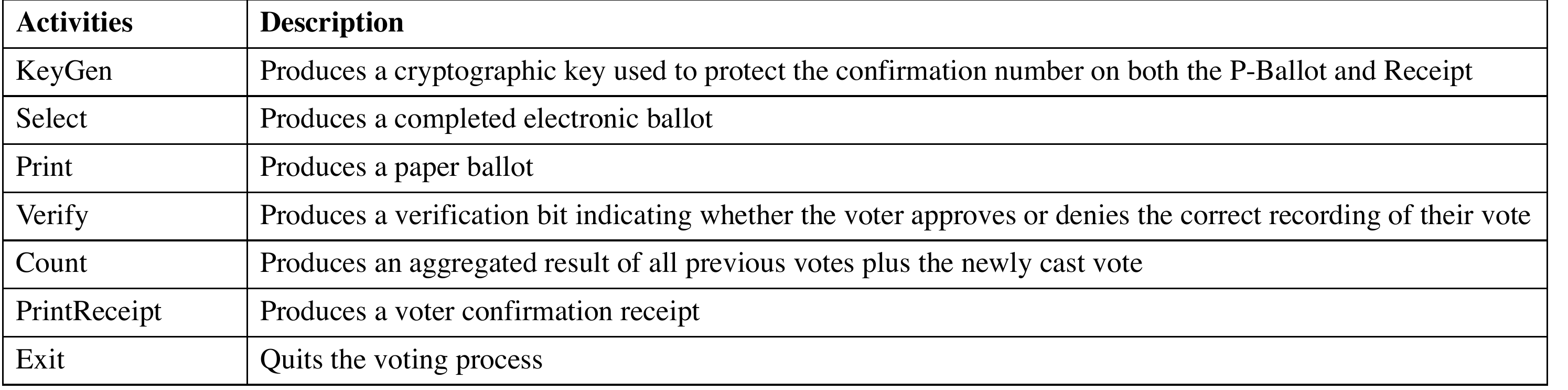}
	\caption{Activities in an ACDC E-voting environment}\label{tab: voter-activities}
	\includegraphics[trim =  .5cm 13.25cm .4cm 2cm,clip, width=\linewidth]{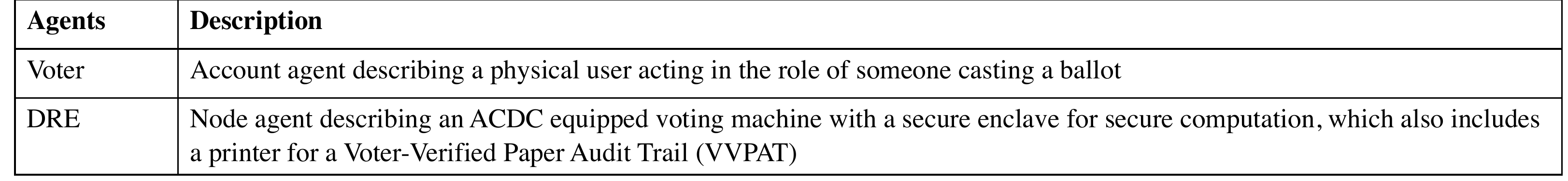}
	\caption{Agents in an ACDC E-voting environment}\label{tab: voter-agents}
\end{table}

\subsection{ACDC E-voting Scenario}

Within an ACDC architecture all voting may take place electronically on ACDC equipped voting machines. For illustration purposes, we assume these voting machines can perform similarly to Direct Recording Electronic (DRE) voting machines with a Voter-Verified Paper Audit Trail (VVPAT)~\cite{wack2005draft}. However, ACDC equipped voting machines perform all computations securely (e.g., in a secure enclave) and the internal data of all capsules is encrypted. Tables~\ref{tab: voter-entities}--\ref{tab: voter-agents} describe the provenance objects in such an ACDC voting network.

In this scenario, a voter's ballot is successfully cast after the following steps: (1)~a voter enters their unique \textit{VoterID} into the ACDC equipped voting machine, (2)~the voting machine invokes a key generation function in which a cryptographic key is generated that will be attributed to the corresponding voter, (3)~the voter will then be presented with an electronic ballot in which they can manually enter their selections, (4)~a paper ballot, containing a cryptographically protected confirmation number, will then be generated and displayed through a viewing glass for a limited amount of time, in which a user can verify whether they approve the recorded selections, (5)~after the user verifies that their vote has been correctly recorded, the machine securely stores the paper ballot for a VVPAT, (6)~the machine then electronically counts the new result by including the newly cast vote, and (7)~the machine then provides a printed receipt to the voter, which includes a cryptographically protected confirmation number that matches the confirmation number of the paper ballot and exclaims that their vote has been counted. The encrypted confirmation number on the receipt provided to the voter can be used at a later date by the voter to ensure that their vote was correctly included in the election result~\cite{bernhard2017public}.

To formalize, let $VM = (\mathcal{E}_{k},\mathcal{E}_{c}, \mathcal{E}_{d}, G_{n}, G_{a},\mathcal{A})$ be a distributed environment of ACDC equipped electronic voting machines where, 
\begin{itemize}
	\item $\mathcal{E}_{k}$ is a finite set of key entities, where each key entity describes a key belonging to either a voter or a voting machine,
	\item $\mathcal{E}_{c}$ is the finite set of contract entities where the possible contracts are \textit{KeyGenContract, SelectContract, PrintContract, VerifyContract, CountContract, PrintReceiptContract}, and \textit{ExitContract},
	\item $\mathcal{E}_{d}$ is a finite set of data entities,
	\item $G_{n}$ is a finite set of node agents, where each node is an ACDC equipped voting machine,
	\item $G_{a}$ is a finite set of account agents, where each account is a physical user of an ACDC equipped voting machine, and
	\item $\mathcal{A}$ is a finite set of activities, where the possible activities are \textit{KeyGen, Select, Print, Verify, Count, PrintReceipt,} and \textit{Exit}.
\end{itemize}

This environment consists of a set of provenance graphs $2^{V}\times2^{E}\times L^{E}$ where $V = \mathcal{E}_{k}\cup \mathcal{E}_{c}\cup \mathcal{E}_{d}\cup G_{n} \cup G_{a} \cup \mathcal{A}$ is the set of all possible vertices, $E \subset V \times V$ is the set of all possible edges, and $L^{E}$ is the set of all possible labeling functions. We assume that in a scenario where a provenance-based enforcement mechanism is tasked with enforcing a provenance policy at a function execution, the mechanism is able to query the provenance record to obtain the relevant provenance graph $(V',E',l') \in 2^{V}\times2^{E}\times L^{E}$. For this particular case study, a mechanism can query the provenance record for all provenance associated with a particular voter. Such an assumption is reasonable because an input-enabled mechanism will be enabled to query the necessary provenance by a voter inputting their \textit{VoterID}; this requirement can be specified by the contract for a specific function.  In this scenario, the provenance graph being evaluated will only contain one account agent, namely the present voter.

\subsection{Voter Fraud Scenarios}

To demonstrate the applicability of ACDC provenance for reasoning about voter fraud in an e-voting context, we consider 2 real scenarios in which voters have committed fraud and present provenance policies that might be enforced by ACDC voting machines to prevent such fraud. Additionally, we present a scenario in which a user may try to manipulate the voting machine and how provenance policies can aid in reasoning about such manipulation. These scenarios include: 1)~a voter attempting to cast multiple votes~\cite{vielmetti2016shorewood,heritagevoterfraud}, 2)~an ineligible voter attempting to cast a vote~\cite{trischitta2013I,heritagevoterfraud}, and 3) a voter attempting to cast multiple votes by exiting the system just before a receipt is printed.\\

\paragraph{Duplicate Voting}

Consider a scenario in which a user, say Alice, is legitimately registered to vote in two states. Although it is not a crime for Alice to be registered in two states, it is a crime, according to state law, for her to cast more than one vote in the same election~\cite{doublevoting}. In this scenario, Alice has intentions on participating in early voting in state 1 and voting on election day in state 2. Because Alice has a legitimate \textit{VoterID} for state 1, her vote will be counted and will result in a provenance record showing that she has cast a legitimate vote. When Alice attempts to vote on election day in state 2, based on her provenance record, the voting machine should not allow her to cast another ballot. The simplest check would be to determine whether Alice has already received a receipt indicating that she has already cast a ballot. To do so, we can express a provenance policy that defines the expected provenance of a printed receipt. This policy can be checked at the execution of the $KeyGen$ activity, as specified by the \textit{KeyGenContract}, when Alice attempts to cast a second ballot. Formally, given a provenance graph $(V',E',l') \in 2^{V}\times 2^{E}\times L^{E}$ that corresponds to all provenance metadata associated with Alice, we can determine whether Alice has been attributed a printed receipt if the following policy $P$ evaluates to true

\begin{flushleft}
	\begin{tabular}{lcp{3in}}
		\small$P(V',E',l')$ & \small $\iff$ &\small ${\exists \varepsilon_{d}, a, g_{a} \in V': ((a, PrintReceiptContract) \in E'}$\\
		&     &\small${\land~l'(a,PrintReceiptContract) = Used)}$\\ 
		&    & \small${\land ((\varepsilon_{d}, a) \in E'\land l'(\varepsilon_{d},a) = WasGeneratedBy)}$\\
		&    & \small$ {\land ((\varepsilon_{d}, PrintReceiptContract) \in E}$\\
		&    &\small${\land~l'(\varepsilon_{d},PrintReceiptContract) = WasDerivedFrom)}$\\ 
		&   & \small${\land ((\varepsilon_{d}, g_{a}) \in E'\land l'(\varepsilon_{d}, g_{a}) = WasAttributedTo))}.$
	\end{tabular}
\end{flushleft}

If the policy evaluates to true over the given provenance graph, then the voting machine can take the necessary actions of preventing Alice from casting a second ballot (e.g., exiting the system).\\

\paragraph{Ineligible Voting}

In the US 2012 election a convicted felon successfully voted in the election, in a state that prohibits convicted felons from voting, by providing false information on the voter registration form~\cite{trischitta2013I}. Consider a scenario in which Bob, who is a convicted felon, falsely indicates that he is not a convicted felon on his voter's registration form and is approved to vote and is provided a legitimate \textit{VoterID}. Because US convicted felon records are public record, this record can be considered as a blacklist of account agents in an ACDC voting network. Although a user may have a valid \textit{VoterID}, voting machines can ensure that they are not acting on behalf of blacklisted account agents. However, to make this determination, Bob will first have to enter his \textit{VoterID} into the voting machine, thereby generating provenance of a voting machine acting on his behalf. When the voting machine invokes the $KeyGen$ function, the function will first use the $KeyGenContract$ to determine how it will process entities. The contract can specify a provenance policy stating that the function should proceed iff the voting machine for which it is associated with is not acting on behalf of a blacklisted account agent. Formally, given Bob's provenance graph $(V',E',l') \in 2^{V}\times 2^{E}\times L^{E}$ we can determine if Bob is a convicted felon if 

\begin{flushleft}
	\begin{tabular}{llcp{3in}}
		\small $\exists G_{a_{blacklist}} \subseteq G_{a}:$ & \small$P(V',E',l')$ & \small$\iff$ & \small$\exists g_{a_{blacklist}} \in G_{a_{blacklist}}:$ \\
		& & & \small$\exists g_{n} \in V': (g_{n}, g_{a_{blacklist}}) \in E'$ \\
		& & & \small$ \land~l'(g_{n},g_{a_{blacklist}}) = ActedOnBehalfOf$.
	\end{tabular}
\end{flushleft} 

If this policy evaluates to true, then it will be known that the voting machine is acting on behalf of a blacklisted user; therefore, this user should not be allowed to cast a vote according to state law. \\

\paragraph{Manipulating an ACDC Voting Machine}

Consider a scenario in which a malicious voter, Mallory,  is aware of the workflow of the voting machine and attempts to manipulate a voting machine into allowing her to vote multiple times by preventing the attribution of a receipt for her vote. In this scenario, Mallory may be able to exit the voting process right after the \textit{Count} function executes but before the \textit{PrintReceipt} function executes. When Mallory attempts to vote again her provenance record will not indicate that she has been attributed a receipt for voting. To detect this scenario, we can specify a policy to detect the execution of each function to determine how far Mallory may have gotten in the voting process. Formally, given a provenance graph $(V',E',l') \in 2^{V}\times 2^{E}\times L^{E}$ we can specify the following policy for the $KeyGen$ function---the other policies can be specified similarly:

\begin{itemize}
\begin{figure}[h]
	\centering
	\begin{minipage}[b]{.49\textwidth}
		\centering
		\includegraphics[trim = 3.5cm 0 3cm 2cm,clip,width=\textwidth]{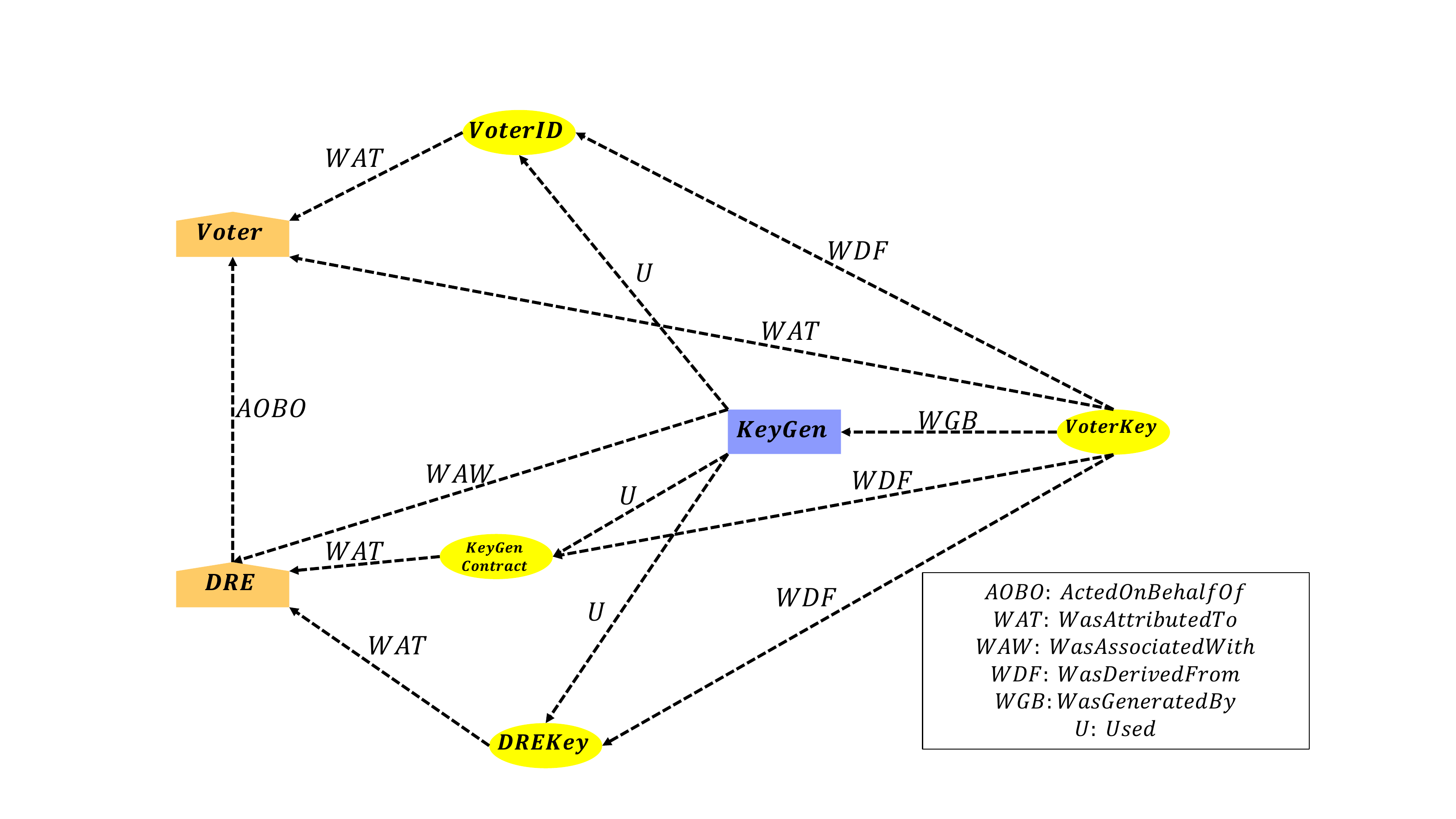}
		\caption{KeyGen provenance event.}\label{fig: kgevent}
	\end{minipage}
	\begin{minipage}[b]{.49\textwidth}
		\centering
		\includegraphics[trim = 10cm 4cm 2cm 6cm,clip,width=\textwidth]{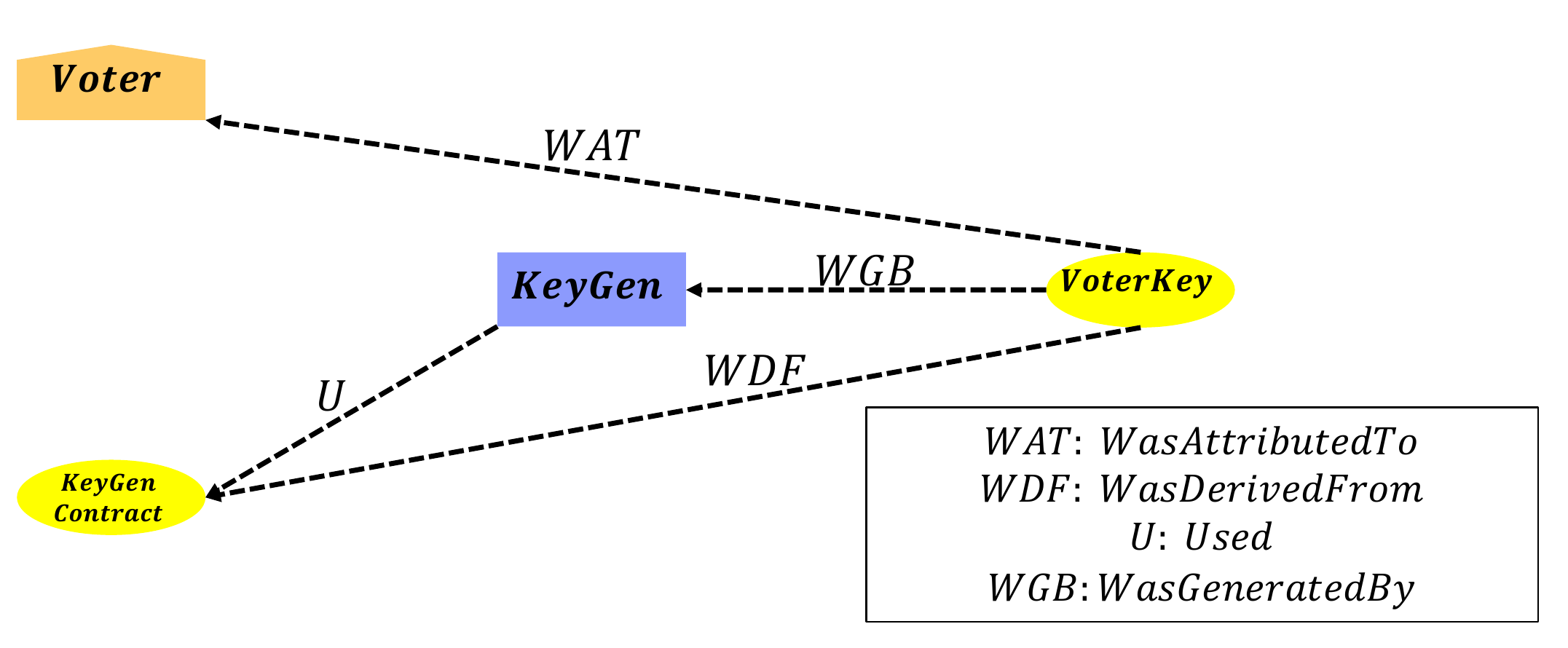}
		\caption{Policy subgraph}\label{fig:kgsub}
	\end{minipage}
\end{figure}
\item KeyGen
\begin{flushleft}
	\begin{tabular}{lcp{3in}}
		\small $P(V',E',l')$ & \small$\iff$ & \small${\exists~\varepsilon_{k}, a, g_{a}\in V': ((a, KeyGenContract) \in E'}$ \\
		& & \small${\land~l'(a,KeyGenContract) = Used)}$ \\
		& & \small${\land~((\varepsilon_{k}, a) \in E'\land l'(\varepsilon_{k},a) = WasGeneratedBy)}$ \\
		& & \small${\land~((\varepsilon_{k}, KeyGenContract) \in E'}$\\
		& & \small${\land~l'(\varepsilon_{k},KeyGenContract) = WasDerivedFrom)}$ \\
		& & \small${\land~((\varepsilon_{k}, g_{a}) \in E'\land l'(\varepsilon_{k}, g_{a}) = WasAttributedTo)}$
	\end{tabular}
\end{flushleft}
\item Select
\begin{flushleft}
\begin{tabular}{lcp{3in}}
\small$P(V',E',l')$ & \small$\iff$ & \small${\exists~\varepsilon_{d}, a, g_{a}\in V': ((a, SelectContract)\in E'}$ \\
& & \small${\land~l'(a,SelectContract) = Used)}$ \\
& & \small${\land~((\varepsilon_{d}, a) \in E'\land l'(\varepsilon_{d},a) = WasGeneratedBy)}$ \\
& & \small${\land~((\varepsilon_{d}, SelectContract) \in E'}$\\
& & \small${\land~l'(\varepsilon_{d},SelectContract) = WasDerivedFrom)}$ \\
& & \small${\land~((\varepsilon_{d}, g_{a}) \in E'\land l'(\varepsilon_{d}, g_{a}) = WasAttributedTo)}$
\end{tabular}
\end{flushleft}
\item Print
\begin{flushleft}
\begin{tabular}{lcp{3in}}
\small$P(V',E',l') $ &\small$\iff$ &\small$\exists~\varepsilon_{d}, a, g_{a}\in V': ((a, PrintContract)\in E'$\\
& & \small${\land~l'(a,PrintContract) = Used) }$ \\
& & \small${\land~((\varepsilon_{d}, a) \in E'\land l'(\varepsilon_{d},a) = WasGeneratedBy)}$ \\
& & \small ${\land~((\varepsilon_{d}, PrintContract) \in E'}$\\
& & \small ${\land~l'(\varepsilon_{d},PrintContract) = WasDerivedFrom)}$ \\
& & \small${\land~((\varepsilon_{d}, g_{a}) \in E'\land l'(\varepsilon_{d}, g_{a}) = WasAttributedTo)}$ \\
\end{tabular}
\end{flushleft}
\item Verify
\begin{flushleft}
\begin{tabular}{lcp{3in}}
\small$P(V',E',l')$ & \small$\iff$ & \small${\exists ~\varepsilon_{d}, a, g_{a}\in V': ((a, VerifyContract)\in E'}$\\
& & \small${\land~ l'(a,VerifyContract) = Used)}$ \\
& & \small${\land~((\varepsilon_{d}, a) \in E'\land l'(\varepsilon_{d},a) = WasGeneratedBy)}$ \\
& & \small${\land~(\varepsilon_{d}, VerifyContract) \in E'}$\\
& & \small${\land~l'(\varepsilon_{d},VerifyContract) = WasDerivedFrom)}$ \\
& & \small${\land~((\varepsilon_{d}, g_{a}) \in E'\land l'(\varepsilon_{d}, g_{a}) = WasAttributedTo}$
\end{tabular}
\end{flushleft}
\item Count
\begin{flushleft}
\begin{tabular}{lcp{3in}}
\small $P(V',E',l') $ & \small $\iff$ & \small${\exists~\varepsilon_{d}, a, g_{n},g_{a}\in V': ((a, CountContract) \in E'}$\\
& & \small${\land~l'(a,CountContract) = Used)}$\\
& & \small$ {\land~((\varepsilon_{d}, a) \in E'\land l'(\varepsilon_{d},a) = WasGeneratedBy)}$ \\
& & \small ${\land~((\varepsilon_{d}, CountContract) \in E'}$\\
& & \small${\land ~l'(\varepsilon_{d},CountContract) = WasDerivedFrom)}$ \\
& & \small ${\land~((\varepsilon_{d}, g_{n}) \in E'\land l'(\varepsilon_{d}, g_{n}) = WasAttributedTo)}$\\
& & \small${\land~((g_n,g_a)\in E'\land l'(g_n,g_a) = ActedOnBehalfOf)}$
\end{tabular}
\end{flushleft}
\end{itemize}

%
%

Informally, such policies can evaluate whether each of the possible contracts were used by activities that generated entities, if so, the generated entities should be derived from the specified contract and attributed to the account agent under consideration or a node agent acting on behalf of the account agent under consideration. Figure's~\ref{fig: kgevent} and~\ref{fig:kgsub} illustrate the $KeyGen$ event and the subgraph specified by the policy, respectively. Similar graphs for each of the other functions and their associated policies can be found in Appendix~\ref{app:events}. These policies can be composed to form a single policy to be evaluated at the \textit{KeyGen} activity whenever a voter attempts to begin the voting process. Because we employ a separation of concerns and specify policies for each functional execution, the mechanism enforcing such policies can determine how far Mallory may have gotten in the voting process by determining which policies fail. In our scenario, since Mallory's provenance record indicates that she completed all steps except for the \textit{PrintReceipt} function, if she attempts to vote on the same machine as her originally counted vote, then the machine can continue its process and print a receipt with a confirmation number based on her \textit{VoterKey}. If Mallory attempts to vote on another machine, then the machine can simply exit, perhaps notifying Mallory to return to the original machine for a receipt.

\subsection{Challenges of Voting Provenance}

Due to the increase of technology used in voting elections where the technology can malfunction~\cite{friedersdorf2018an}, is possibly vulnerable to attacks~\cite{appel2009new}, and may be hacked~\cite{bannet2004hack},  it is important to be able to verify the trustworthiness of results reported by voting machines. Data provenance collection is one viable solution to ensure trustworthy results. However, in a democratic election it is important to only reveal the final result of the election while keeping individual votes secret. Auditing the provenance record of a DRE voting machine in a traditional provenance architecture can reveal the results of individual ballots and can attribute ballots to specific voters. 

Prior work has examined protection mechanisms for provenance storage systems in which the leakage of the provenance record is potentially more sensitive than the leakage of the data for which the provenance corresponds (e.g.,~\cite{bates2013towards,braun2008securing}). However, such solutions are system-centric, relying on protection mechanisms of the storage system. If the system is breached by an unauthorized agent, the provenance record may be exposed. Therefore, the security of the provenance record relies on the strength of security placed on the physical storage system.

We argue that a data-centric approach is more suitable and may provide better security guarantees in scenarios where both the data and provenance record of such data can reveal sensitive information. Analyzing provenance records in an ACDC e-voting network, where all data capsules contain encrypted data, does not suffer from the drawbacks of analyzing provenance records in a traditional system-centric architecture because an ACDC provenance record is a causal record of named encrypted data rather than a causal record of named plaintext data. Therefore, the only information that may be revealed by an ACDC voting provenance record is that a specific user cast a vote but not what or who the particular user voted for. We do not consider revealing that a particular user cast a vote as a limitation of this architecture because this fact is inherent to any voting system in practice.

\section{Related Work}\label{sec: background}
Several frameworks have been proposed for analyzing provenance metadata but do so reactively and in retrospect, relying on either human analysis or the use of automated tools that may rely on machine learning techniques to characterize provenance graphs. Reactive security has benefits in areas such as identifying the root cause of an attack~\cite{lee2013high} and security auditing to ensure compliance with company policies~\cite{pasquier2018data}. While useful, these security practices do not actively prevent security mishaps. Proactive security practices should also be used in conjunction with reactive security practices. However, because proactive security policies are specified with the intent of being enforced, such policies must be based on precise and unambiguous reasoning instead of human intuition. Relevant to this work is proactive reasoning about data provenance, which has received little attention in the literature. 

Much work related to data provenance has focused in the areas of provenance collection (e.g.,~\cite{liang2017provchain}) and secure storage of provenance metadata (e.g.,~\cite{liang2017towards}). Both of these areas are foundational to provenance-aware systems; however, in the context of security, it is equally important to continually analyze provenance metadata at runtime to gain insight into and maintain a computing environment's overall security posture.

Due to the large amounts of data that provenance collection systems can capture, relying on human analysis is impractical and error prone~\cite{hassan2018towards}. Automated tools aim to simplify and make the analysis of provenance metadata more efficient; however, many do so at a loss in precision. Huynh et al.~\cite{huynh2018provenance} present an automated analysis technique that relies on network analysis and machine learning techniques, it is shown that their analysis technique is able to classify provenance graphs into predetermined categories with high accuracy.  FRAPpuccino~\cite{han2017frappuccino} is a provenance-based intrusion detection framework that aims to distinguish benign from anomalous behavior using a machine learning approach. Although machine learning techniques improve the efficiency with which provenance graphs can be analyzed, in high security contexts, such techniques have at least two drawbacks: (1) the classification categories do not provide well-defined properties of the graphs, and (2) the classification categories cannot provide formal guarantees about data due to the possibility of false positives and false negatives.

CamQuery~\cite{pasquier2018runtime} is a framework for the runtime analysis of whole system provenance. Because analysis takes place at runtime, the framework takes a proactive approach to policy specification over provenance metadata by expressing policies in a programmable graph processing framework inspired by GraphChi~\cite{kyrola2012graphchi} and GraphX~\cite{gonzalez2014graphx}. Our approach differs from CamQuery in that we present a formal approach for reasoning about provenance policies in a distributed environment, which is based on a mathematical semantics of provenance graphs.

Lemay et al.~\cite{lemay2017automated} present a framework for automated analysis of provenance by using graph grammars as a way to characterize provenance graphs. However, because the class of graphs parseable by such grammar is restricted to regular grammars, precision is lost and some graphs become parseable that the analyst may not intend to be; therefore, this approach is not amenable to security policy specification in which the policy must be precise and unambiguous. 

Park et al.~\cite{park2012provenance}, present a model for provenance-based access control in which policies are specified using propositional logic as an underlying formalism. This approach can provide formal guarantees about data that conforms to the policy. However, the approach presented in~\cite{park2012provenance} is specific to the access-control domain. In this paper, we have provided a more general and expressive framework for reasoning about provenance policies in a distributed, data-centric environment by using predicate logic as an underlying formalism.

\section{Conclusion and Future Work}\label{sec:conc}

In summary, this paper presented a new data-centric paradigm that provides capabilities for rigorous provenance analysis over distributed systems. A formal approach for reasoning about, and the proactive specification of, provenance policies was introduced. Additionally. we provided a case study that examined the provenance policies necessary to ensure integrity of an ACDC-equipped electronic voting system without sacrificing capabilities for post-factum auditing that traditional provenance techniques provide. We believe that the migration from the current server-centric security paradigm is key to not only enabling the collection of coarsely-grained provenance that is suitable for proactive policy evaluation, but also defends against catastrophic compromises of data records within a given system. In this regard, there are two primary directions for future work stemming from this initial policy design and evaluation. First, the expansion of the ACDC framework. Securing data as a first-class citizen is an approach that has a myriad of benefits that prevent many of the pitfalls that have led to catastrophic data breaches in systems today. Second, there is independent advancement of  provenance policies in the Function as a Service (FaaS) execution model. Such an expansion could enable clients of services such as AWS lambda to untangle the currently inscrutable chain of custody for inputs and products used in FaaS-style execution. This may entail the introduction of a distributed truncation-resistant store and provenance hooks into FaaS job specifications, but could be handled entirely on the clients' end.

\bibliography{related-work}{}

\begin{thebibliography}{10}
\providecommand{\url}[1]{\texttt{#1}}
\providecommand{\urlprefix}{URL }
\providecommand{\doi}[1]{https://doi.org/#1}

\bibitem{heritagevoterfraud}
A sampling of election fraud cases from across the country.
  \url{https://www.heritage.org/sites/default/files/voterfraud_download/VoterFraudCases_5.pdf},
  accessed: 2020--01--10

\bibitem{doublevoting}
Double voting.
  \url{https://www.ncsl.org/research/elections-and-campaigns/double-voting.aspx}
  (2018), accessed: 2020--01--10

\bibitem{appel2009new}
Appel, A.W., Ginsburg, M., Hursti, H., Kernighan, B.W., Richards, C.D., Tan,
  G., Venetis, P.: {The New Jersey Voting-machine Lawsuit and the AVC Advantage
  DRE Voting Machine.} Electronic Voting Technology Workshop/Workshop on
  Trustworthy Elections.  (2009)

\bibitem{bannet2004hack}
Bannet, J., Price, D.W., Rudys, A., Singer, J., Wallach, D.S.: {Hack-a-vote:
  Security Issues with Electronic Voting Systems}. IEEE Security \& Privacy
  \textbf{2}(1),  32--37 (2004)

\bibitem{bates2013towards}
Bates, A., Mood, B., Valafar, M., Butler, K.: Towards secure provenance-based
  access control in cloud environments. In: Proceedings of the third ACM
  conference on Data and application security and privacy. pp. 277--284. ACM
  (2013)

\bibitem{w3c-prov-dm}
Belhajjame, K., B'Far, R., Cheney, J., Coppens, S., Cresswell, S., Gil, Y.,
  Groth, P., Klyne, G., Lebo, T., McCusker, J., Miles, S., Myers, J., Sahoo,
  S., Tilmes, C.: {PROV-DM: The PROV Data Model}. Tech. rep. (2012),
  \url{http://www.w3.org/TR/prov-dm/}

\bibitem{bernhard2017public}
Bernhard, M., Benaloh, J., Halderman, J.A., Rivest, R.L., Ryan, P.Y., Stark,
  P.B., Teague, V., Vora, P.L., Wallach, D.S.: {Public Evidence from Secret
  Ballots}. In: International Joint Conference on Electronic Voting. pp.
  84--109. Springer (2017)

\bibitem{braun2008securing}
Braun, U.J., Shinnar, A., Seltzer, M.I.: {Securing Provenance}. In:
  {Proceedings of the 3rd USENIX Workshop on Hot Topics in Security} (2008)

\bibitem{cassidy2018voting}
Cassidy, C.A., Long, C.: Voting officials under scrutiny amid heavy election
  turnout. \url{https://apnews.com/8af093ef14954d3293fae718c37f3eb3} (2018),
  accessed: 2020--01--10

\bibitem{chase2007multi}
Chase, M.: Multi-authority attribute based encryption. In: Theory of
  cryptography conference. pp. 515--534. Springer (2007)

\bibitem{friedersdorf2018an}
Friedersdorf, C.: An embarrassment of glitches: A wealthy country should be
  able to conduct a national election with fewer problems than the united
  states experiences in the 2018 midterms.
  \url{https://www.theatlantic.com/ideas/archive/2018/11/voting-machines/575044/}
  (2018), accessed: 2020-01--10

\bibitem{gonzalez2014graphx}
Gonzalez, J.E., Xin, R.S., Dave, A., Crankshaw, D., Franklin, M.J., Stoica, I.:
  Graphx: Graph processing in a distributed dataflow framework. In: {11th
  USENIX Symposium on Operating Systems Design and Implementation}. pp.
  599--613 (2014)

\bibitem{han2017frappuccino}
Han, X., Pasquier, T., Ranjan, T., Goldstein, M., Seltzer, M.: Frappuccino:
  Fault-detection through runtime analysis of provenance. In: Workshop on Hot
  Topics in Cloud Computing (2017)

\bibitem{hassan2018towards}
Hassan, W.U., Aguse, L., Aguse, N., Bates, A., Moyer, T.: Towards scalable
  cluster auditing through grammatical inference over provenance graphs. In:
  Network and Distributed Systems Security Symposium (2018)

\bibitem{huynh2018provenance}
Huynh, T.D., Ebden, M., Fischer, J., Roberts, S., Moreau, L.: Provenance
  network analytics. Data Mining and Knowledge Discovery  \textbf{32}(3),
  708--735 (2018)

\bibitem{Jacobson:2009:NNC:1658939.1658941}
Jacobson, V., Smetters, D.K., Thornton, J.D., Plass, M.F., Briggs, N.H.,
  Braynard, R.L.: {Networking Named Content}. In: Proceedings of the 5th
  International Conference on Emerging Networking Experiments and Technologies.
  pp. 1--12 (2009)

\bibitem{kyrola2012graphchi}
Kyrola, A., Blelloch, G., Guestrin, C.: {GraphChi: Large-Scale Graph
  Computation on Just a PC}. In: {10th USENIX Symposium on Operating Systems
  Design and Implementation}. pp. 31--46 (2012)

\bibitem{lee2013high}
Lee, K.H., Zhang, X., Xu, D.: High accuracy attack provenance via binary-based
  execution partition. In: {Network and Distributed System Security Symposium}
  (2013)

\bibitem{lemay2017automated}
Lemay, M., Hassan, W.U., Moyer, T., Schear, N., Smith, W.: Automated provenance
  analytics: a regular grammar based approach with applications in security.
  In: {9th USENIX Workshop on the Theory and Practice of Provenance } (2017)

\bibitem{liang2017provchain}
Liang, X., Shetty, S., Tosh, D., Kamhoua, C., Kwiat, K., Njilla, L.: Provchain:
  A blockchain-based data provenance architecture in cloud environment with
  enhanced privacy and availability. In: Proceedings of the International
  Symposium on Cluster, Cloud and Grid Computing. pp. 468--477. IEEE Press
  (2017)

\bibitem{liang2017towards}
Liang, X., Zhao, J., Shetty, S., Li, D.: Towards data assurance and resilience
  in iot using blockchain. In: IEEE Military Communications Conference. pp.
  261--266. IEEE (2017)

\bibitem{park2012provenance}
Park, J., Nguyen, D., Sandhu, R.: A provenance-based access control model. In:
  International Conference on Privacy, Security and Trust. pp. 137--144. IEEE
  (2012)

\bibitem{pasquier2018runtime}
Pasquier, T., Han, X., Moyer, T., Bates, A., Hermant, O., Eyers, D., Bacon, J.,
  Seltzer, M.: Runtime analysis of whole-system provenance. In: Proceedings of
  the 2018 ACM SIGSAC Conference on Computer and Communications Security. pp.
  1601--1616. ACM (2018)

\bibitem{pasquier2018data}
Pasquier, T., Singh, J., Powles, J., Eyers, D., Seltzer, M., Bacon, J.: Data
  provenance to audit compliance with privacy policy in the internet of things.
  Personal and Ubiquitous Computing  \textbf{22}(2),  333--344 (2018)

\bibitem{trischitta2013I}
Trischitta, L.: {`I voted early' sticker leads to arrest, fraud charges}.
  \url{https://www.sun-sentinel.com/news/fl-xpm-2013-02-22-fl-felon-voter-fraud-pompano-20130222-story.html}
  (2013), accessed: 2020--01--10

\bibitem{vielmetti2016shorewood}
Vielmetti, B.: Shorewood man sentenced to jail for multiple votes in several
  elections.
  \url{https://archive.jsonline.com/news/crime/shorewood-man-sentenced-to-jail-for-multiple-votes-in-several-elections-b99677321z1-370317801.html},
  accessed: 2020--01--10

\bibitem{wack2005draft}
Wack, J.P.: {Draft Standard for Voter Verified Paper Audit Trails in DRE Voting
  Systems (DRE-VVPAT): Supplement to the 2002 Voting Systems Standard}.
  \url{https://www.nist.gov/system/files/documents/itl/vote/VVPAT-Addendum-jpw-3-2-051.pdf}
  (2005), accessed: 2020--01--10

\end{thebibliography}
\bibliographystyle{splncs}
\newpage
\appendix
\section{Provenance Graphs of Individual Case Study Events}\label{app:events}
\begin{figure}[h]
\centering
\begin{minipage}[b]{.49\textwidth}
\centering
\includegraphics[trim = 2.5cm 0 0 0,clip,width=\textwidth]{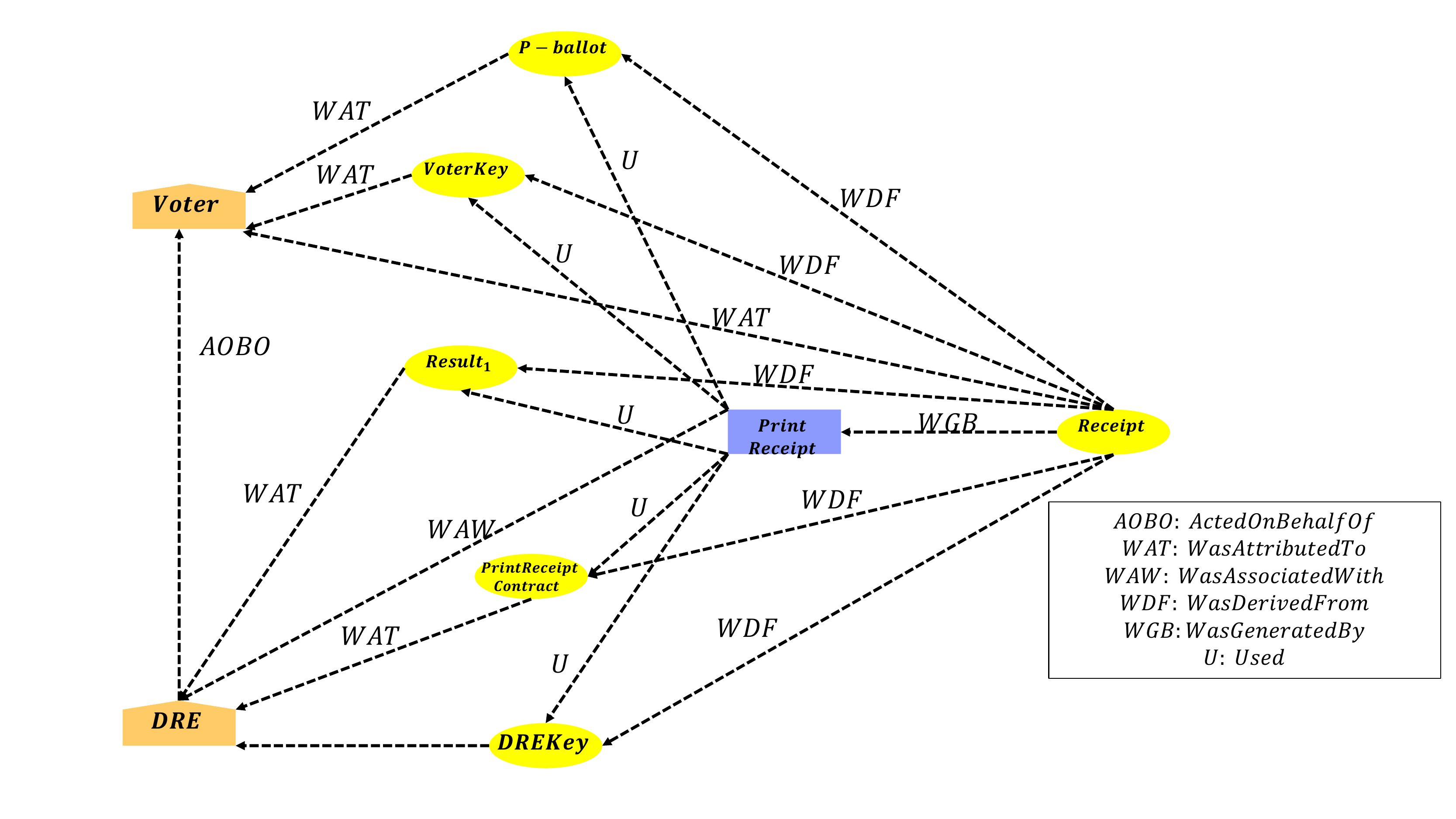}
\caption{PrintReceipt provenance event}
\end{minipage}
\begin{minipage}[b]{.49\textwidth}
\includegraphics[trim = 10.5cm 3.5cm 1cm 6cm,clip,width=\textwidth]{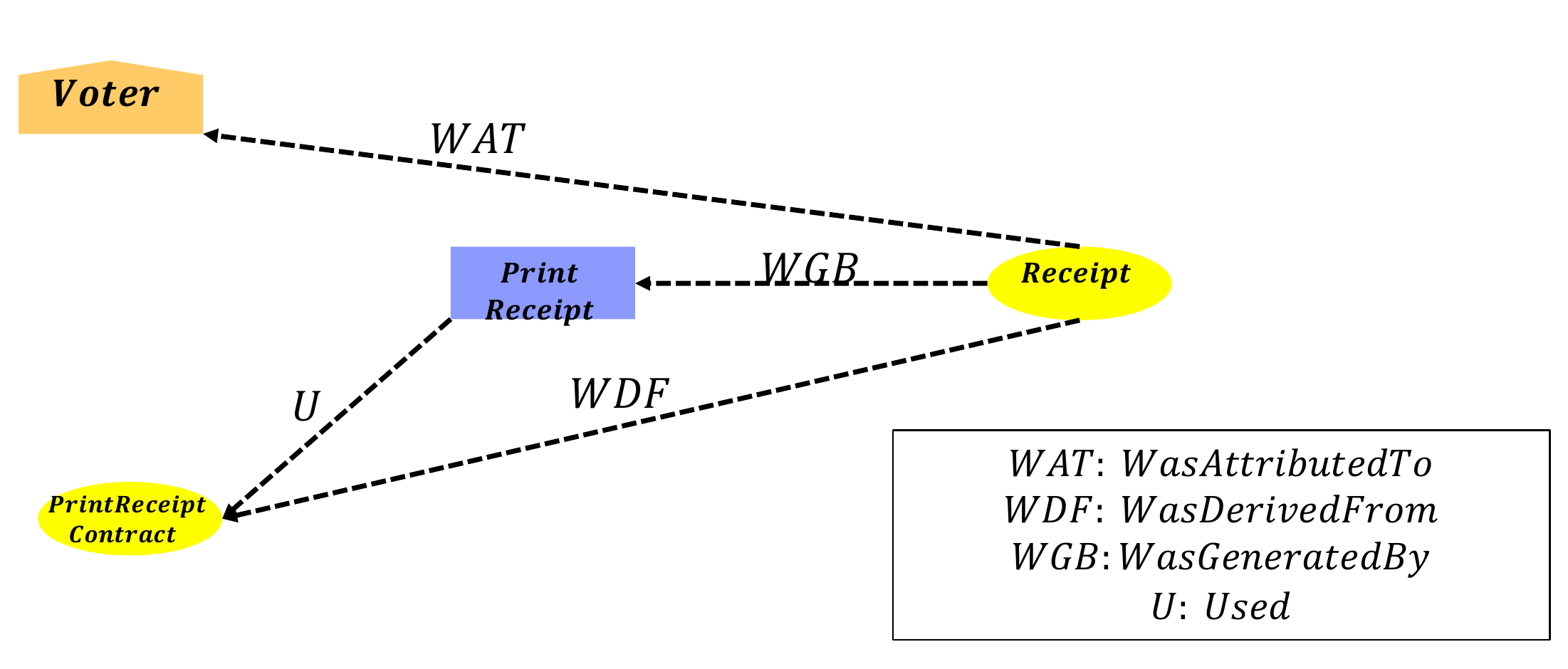}
\caption{Policy subgraph }
\end{minipage} 
\end{figure}

\begin{figure}[h]
\centering
\begin{minipage}[b]{.49\textwidth}
\centering
\includegraphics[trim = 3.5cm 0 0 2cm,clip,width=\linewidth]{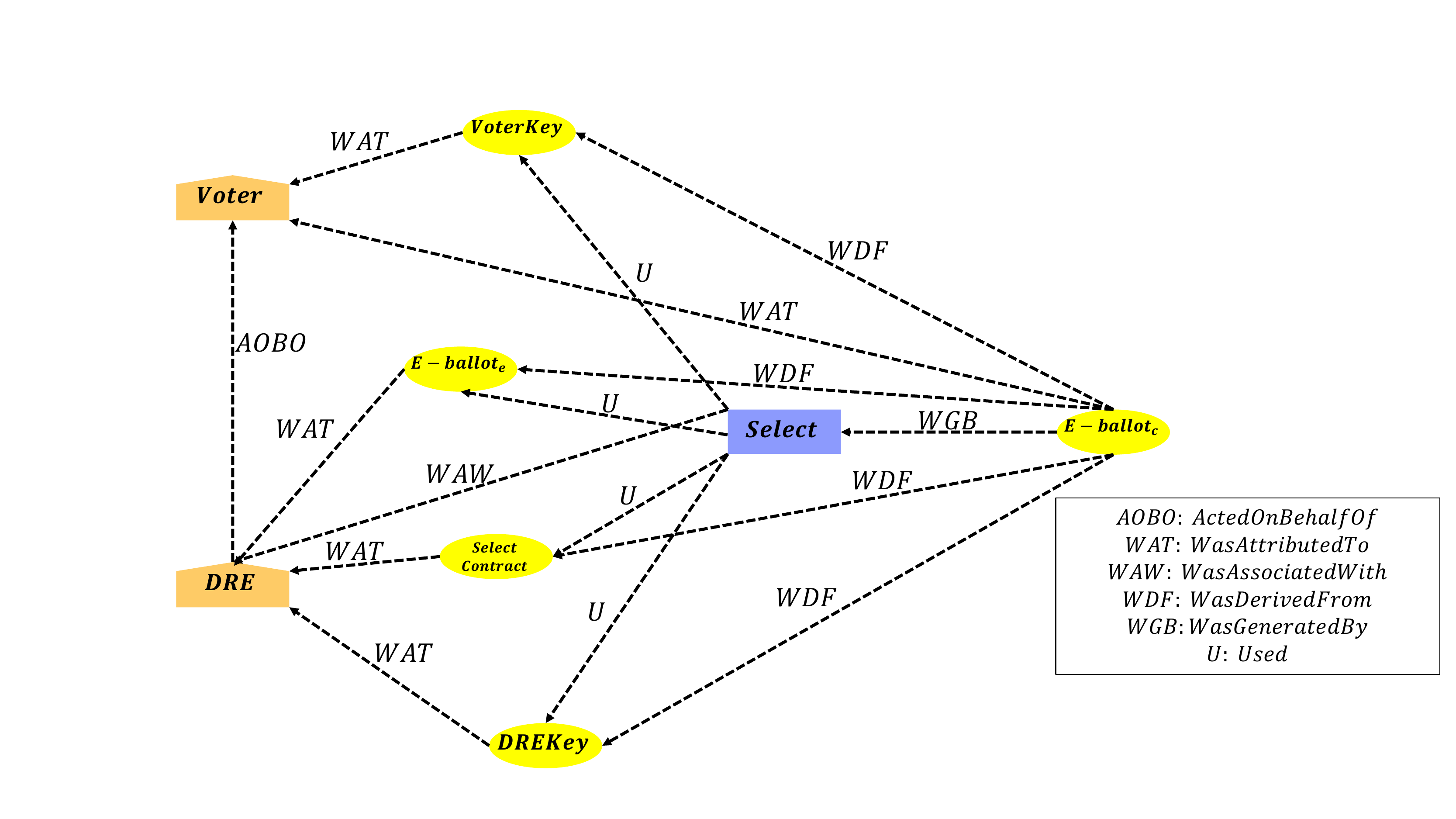}
\caption{Select provenance event.}\label{fig: selectevent}
\end{minipage}
\begin{minipage}[b]{.49\textwidth}
\centering
\includegraphics[trim= 10cm 4.5cm .75cm 6.5cm,clip,width=\textwidth]{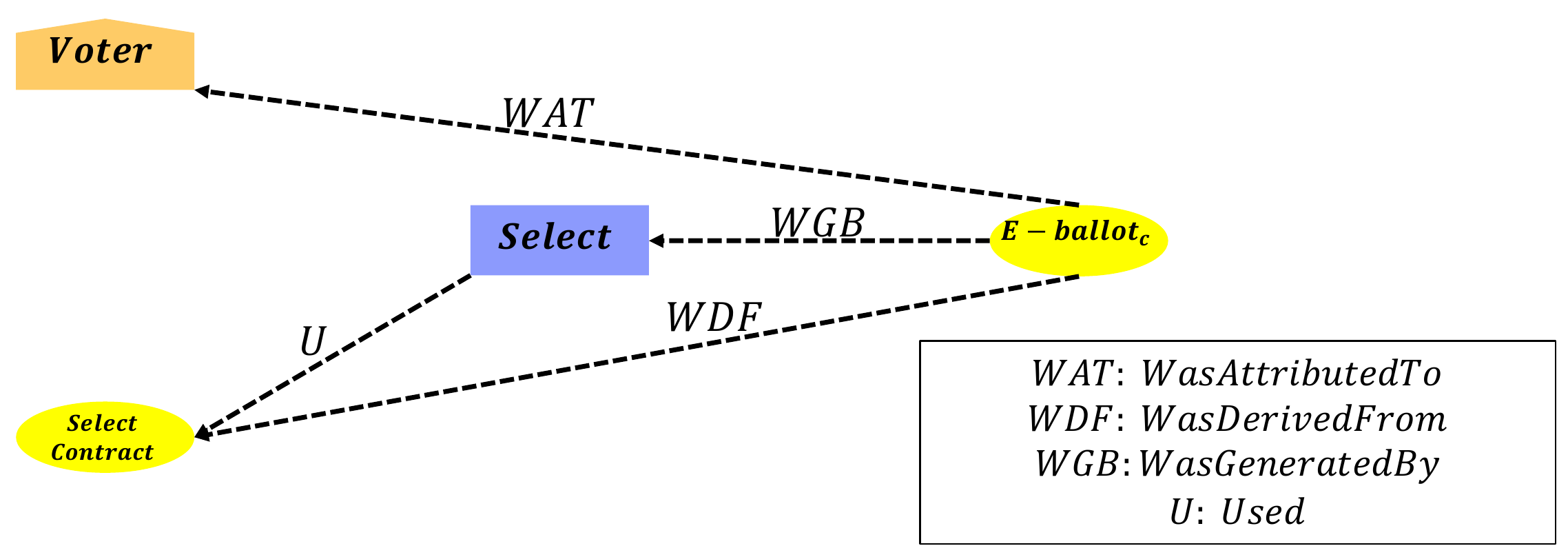}
\caption{Policy subgraph}
\end{minipage}
\end{figure}
\begin{figure}[h]
\centering
\begin{minipage}[b]{.49\textwidth}
\centering
\includegraphics[trim = 3.5cm 0 0 2cm, clip,width=\textwidth]{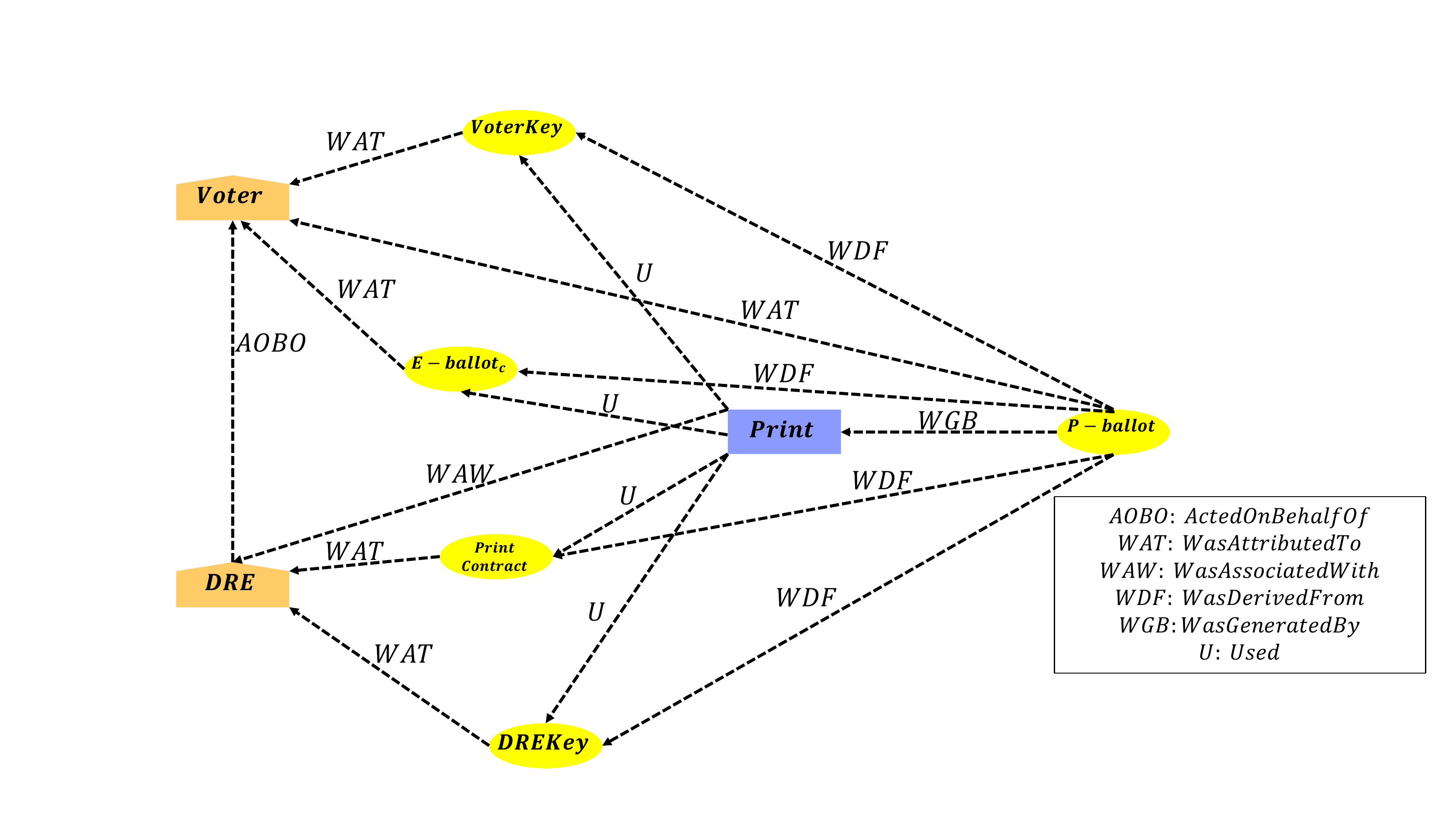}
\caption{Print provenance event}
\end{minipage}
\begin{minipage}[b]{.49\textwidth}
\centering
\includegraphics[trim = 10cm 4.5cm .75cm 6.5cm,clip,width=\textwidth]{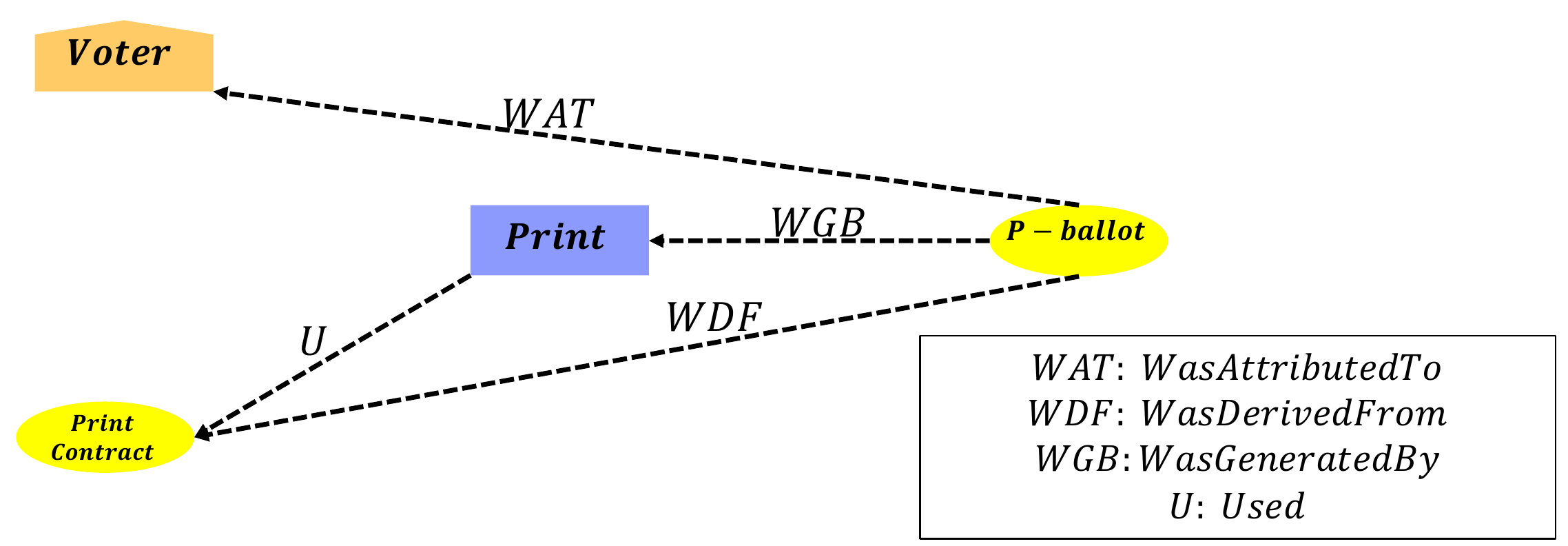}
\caption{Policy subgraph}
\end{minipage}
\end{figure}
%
\begin{figure}[h]
\centering
\begin{minipage}[b]{.49\textwidth}
\centering
\includegraphics[trim = 3.5cm 0 0 2cm, clip,width=\textwidth]{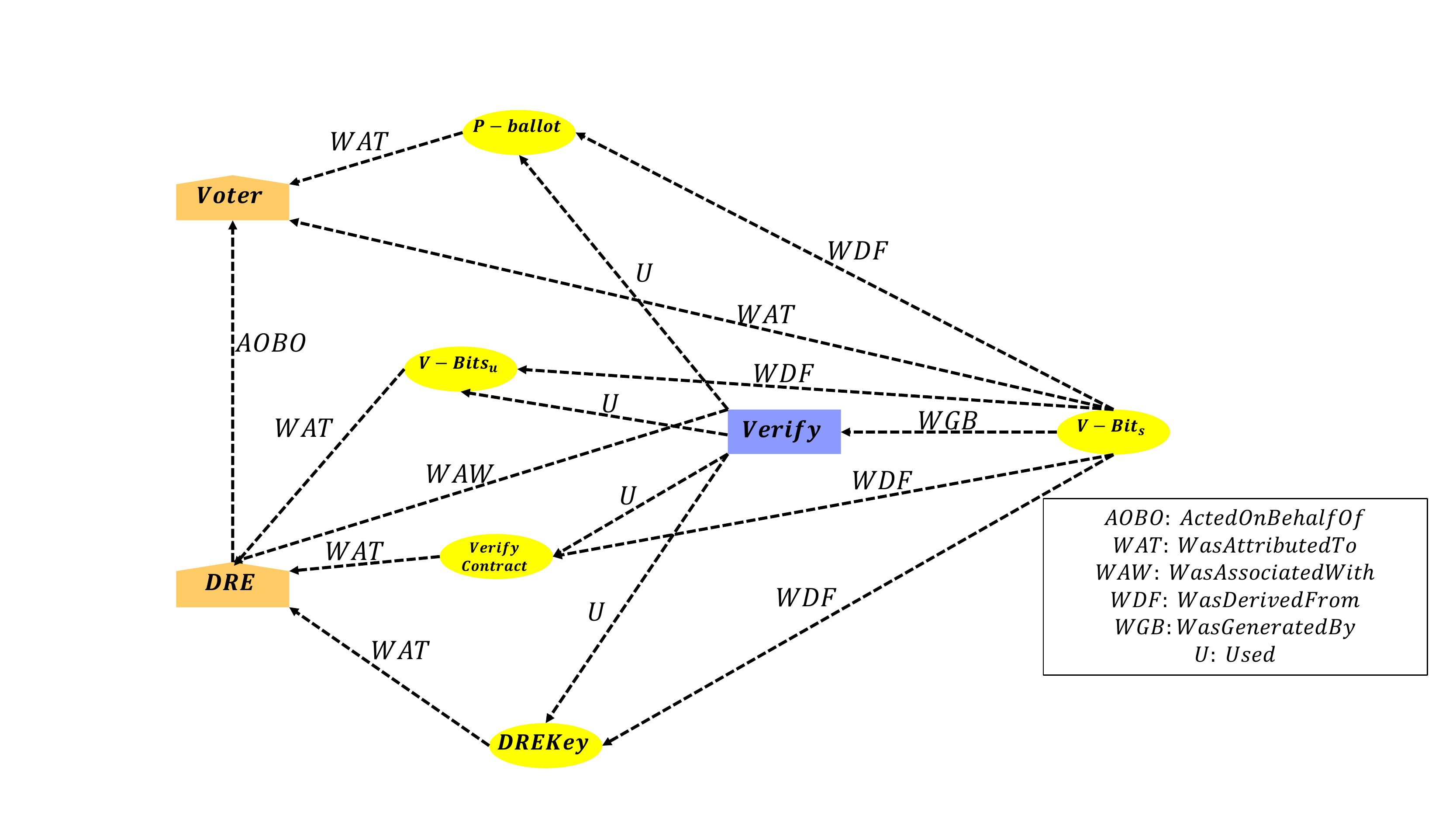}
\caption{Verify provenance event}
\end{minipage}
\begin{minipage}[b]{.49\textwidth}
\centering
\includegraphics[trim = 10cm 4.5cm .75cm 6.5cm,clip,width=\textwidth]{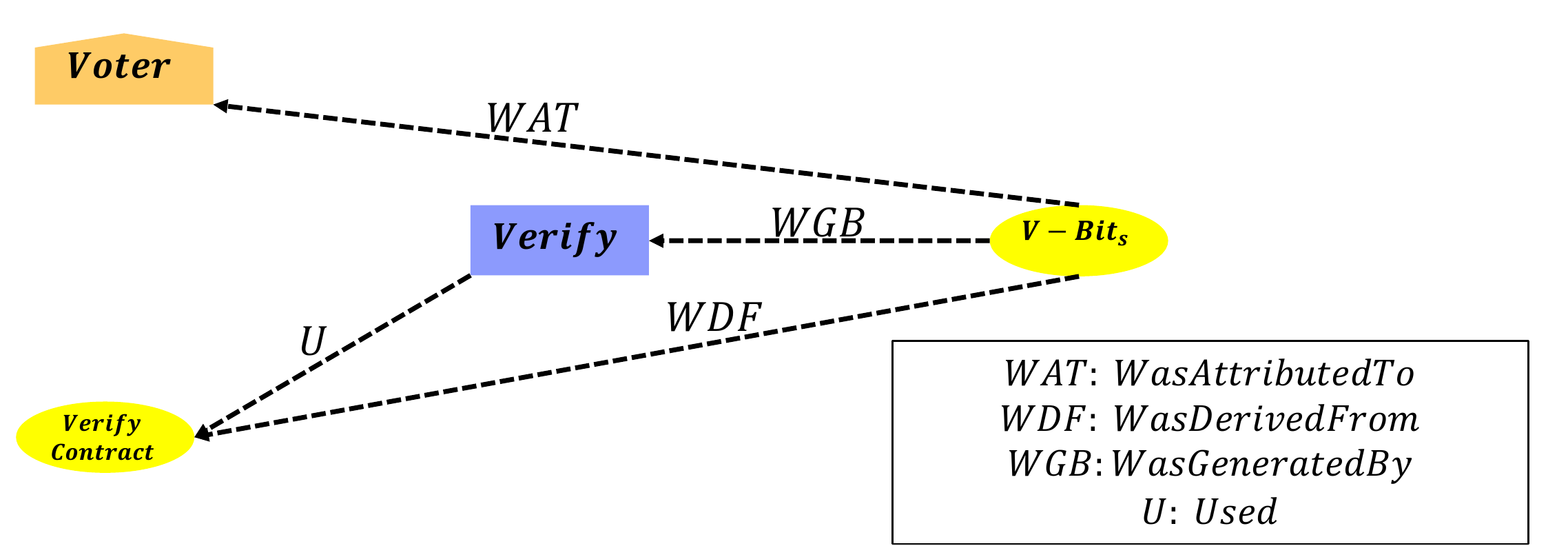}
\caption{Policy subgraph}
\end{minipage}
\end{figure}

\begin{figure}[h]
\centering
\begin{minipage}[b]{.49\textwidth}
\centering
\includegraphics[trim = 3cm 0 .65cm .7cm,clip,width=\textwidth]{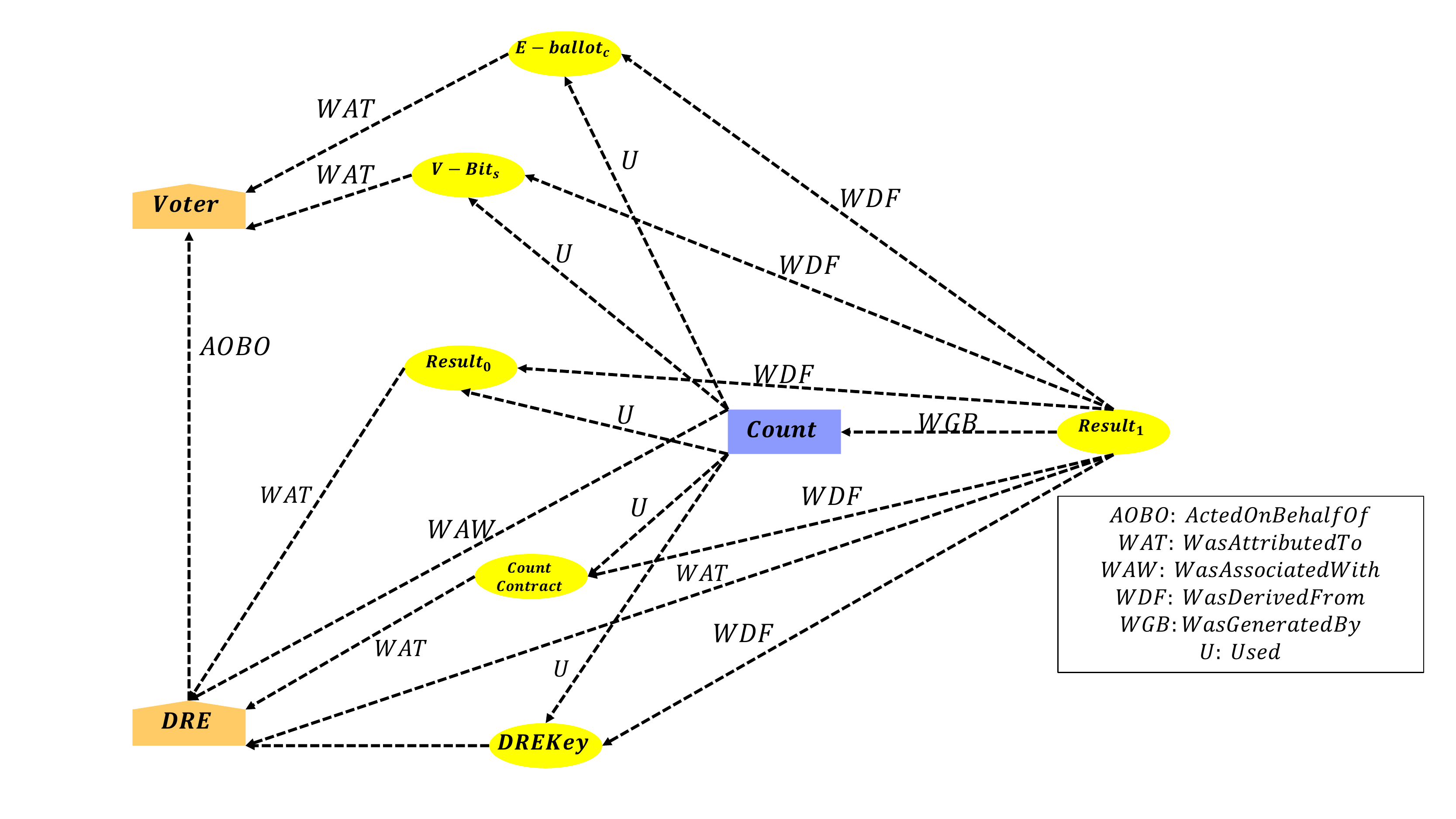}
\caption{Count provenance event}
\end{minipage}
\begin{minipage}[b]{.49\textwidth}
\centering
\includegraphics[trim = 7cm 1cm 3.5cm 9cm,clip,width=\textwidth]{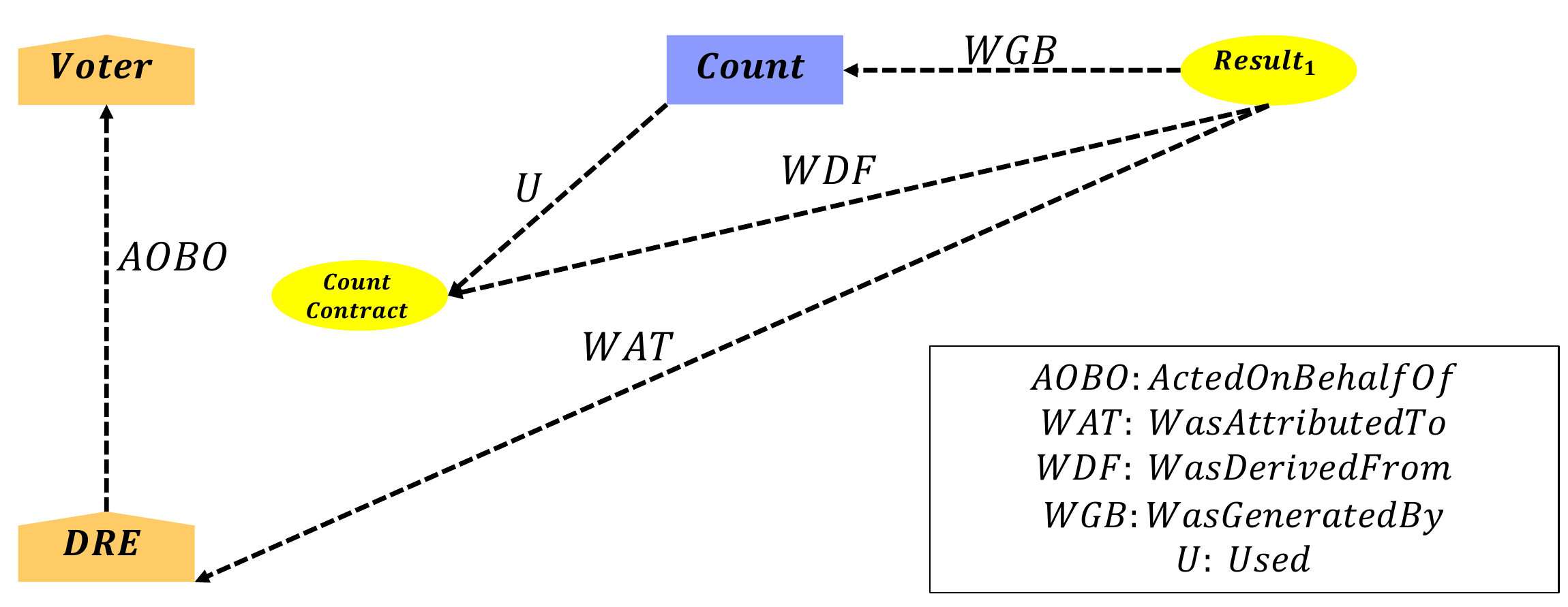}
\caption{Policy subgraph}
\end{minipage}
\end{figure}

%
\end{document}